\documentclass[reprint,
nofootinbib,
 amsmath,amssymb,
 aps,
floatfix,
]{revtex4-2}
\usepackage{empheq}
\usepackage{dcolumn}% Align table columns on decimal point
\usepackage{bm}% bold math
\usepackage{hyperref}% add hypertext capabilities
\usepackage[mathlines]{lineno}% Enable numbering of text and display math
\usepackage{float}

\usepackage{enumitem}
\usepackage{braket}
\usepackage{tikz}
\usepackage{graphicx}% Include figure files

\usepackage{dcolumn}% Align table columns on decimal point
\usepackage{bm}% bold math
\usepackage{hyperref}% add hypertext capabilities
\usepackage[mathlines]{lineno}% Enable numbering of text and display math
\usepackage{mathtools}
\usepackage[dvipsnames]{xcolor}

\DeclareRobustCommand{\Moustafa}[1]{{#1}}

\numberwithin{equation}{section}

\newcommand{\MAGIC}{\textsc{magic}}

\makeatletter

\renewcommand*{\p@subsection}{}

\renewcommand*{\p@subsubsection}{}
\makeatother
\begin{document}

\title{Metric--affine gravity with dynamical chronology protection}

 \author{Moustafa Ismail}
\email{MoustafaKamel.Ismail@unh.edu}%Lines break automatically or can be forced with \\
\author{David Mattingly}%
 \email{David.Mattingly@unh.edu}
\affiliation{University of New Hampshire, Durham, New Hampshire 03824, USA}

\date{\today}% It is always \today, today,
             %  but any date may be explicitly specified

\begin{abstract}
Modified theories of gravity often introduce geometric structure beyond general relativity to address unresolved problems in the gravitational sector without invoking ad hoc matter fields. Mimetic gravity, for example, generates an effective cosmological dark sector by isolating the conformal mode of the metric, while Ho\v{r}ava--Lifshitz gravity attains power-counting renormalizability by endowing spacetime with a preferred dynamical foliation. Although chronology protection was not the original motivation for either theory, both can realize it classically through stable causality. This suggests that chronology protection itself may be elevated from a derived property to a guiding principle for constructing modified gravitational theories, especially if its implementation at the quantum-gravitational level leaves infrared imprints in the effective action. Motivated by this possibility, we introduce a toy metric--affine gravity model whose chronology-protecting sector is purely geometric. The model breaks projective invariance and selects a timelike, closed non-metricity one-form. In the globally exact sector, this one-form defines a time function, realizing stable causality through affine geometry. On the matter-free, spatially flat FLRW branch \Moustafa{with non-compact slices, $\Lambda=0$, and positive chronology-sector density}, the \Moustafa{local quadratic} gravitational \Moustafa{spectrum} contains the two tensor modes of GR and a single scalar mode. \Moustafa{The scalar has positive kinetic energy but a gradient instability, $c_s^2=-2$, requiring stabilization before a viable cosmological application can be established.} We investigate the classical cosmological effects and scalar-mode dynamics of this purely geometric chronology-protecting extension of general relativity.
\end{abstract}

%\keywords{Suggested keywords}%Use showkeys class option if keyword
                              %display desired
\maketitle

%\tableofcontents

\section{\label{sec:Introduction}Introduction}

A physically viable theory is expected to admit a well-posed notion of time evolution. Given suitable initial data on an initial hypersurface, the theory determines the subsequent evolution of observables uniquely. This, however, presupposes a globally consistent notion of temporal ordering. If no such ordering exists, spacetime may admit a \emph{closed timelike curve} (CTC), along which an observer can return to an earlier event on their own worldline. The existence of CTCs threatens both causality and predictability. Once causal curves can close, the usual initial-value formulation ceases to be globally well defined, and familiar paradoxes may arise, such as the grandfather paradox, in which a time traveler prevents their own existence by preventing their grandparents from meeting.

Classical general relativity (GR) admits exact solutions with CTCs. Standard examples include the G{\"o}del universe~\cite{Godel}, the Tipler cylinder~\cite{TiplerTimeMachine}, Gott's time-machine spacetime~\cite{Gott}, traversable wormhole constructions~\cite{WormhoelTimeMachine}, and regions of the maximally extended Kerr geometry~\cite{Kerr}. One possible resolution is to keep such geometries but restrict the allowed histories on them. Novikov's self-consistency principle, for example, permits only globally self-consistent evolutions on spacetimes containing CTCs~\cite{Novikov,Novikov1989}. While this avoids contradiction, it does so at the cost of severely restricting the allowed initial data~\cite{CauchyCTC,Novikov1989,WormhoelTimeMachine,GenericTimeMachine}. A more conservative viewpoint is that CTCs are simply unphysical and should be excluded by the fundamental laws of nature. This is Hawking's \emph{chronology protection conjecture}, namely that the laws of physics---quantum gravity in particular---do not allow the appearance of CTCs~\cite{CPC,VisserCPC,VisserCPC2}.

There are good reasons to suspect that quantum effects become relevant when one attempts to form a time machine. Hawking argued that vacuum polarization near a chronology horizon should obstruct the formation of CTCs~\cite{CPC}. Kay, Radzikowski, and Wald later proved that, for spacetimes with a compactly generated Cauchy horizon, the two-point function of a free quantum field generically fails to be Hadamard on the horizon~\cite{KRW}. Because the renormalized energy-momentum tensor in semiclassical gravity is defined only for Hadamard states, the semiclassical Einstein equation ceases to be well defined at such points, even before a fully developed acausal region is reached~\cite{KRW,VisserCPC,VisserCPC2}.

A closely related lesson arises in black-hole interiors. In Kerr and Reissner--Nordstr\"om spacetimes, the inner horizon is a Cauchy horizon, and perturbations falling inward experience an infinite blueshift there. Classical backreaction then leads to mass-inflation instability, strongly suggesting that the na{\"i}ve smooth extension beyond the Cauchy horizon is non-generic~\cite{PoissonIsrael,OriMassInflation}. Both chronology horizons and inner Cauchy horizons therefore point toward the same qualitative conclusion: whenever causality or predictability is endangered, the low-energy description based on classical or semiclassical gravity is insufficient; a complete quantum theory of gravity is required.

If chronology protection is true and enforced by \emph{some} ultraviolet mechanism in the gravitational sector, it may leave behind a low-energy imprint in the effective gravitational theory.
This logic is familiar from broader searches for infrared echoes of ultraviolet structure, including Lorentz violation, fifth forces, extra dimensions, and low-energy supersymmetry. Since pure GR admits exact solutions with CTCs, any low-energy remnant of chronology protection must depart from GR in some way. For example, Causal Dynamical Triangulations exhibits infrared departures from GR~\cite{CDT,LCDT,PHorava,SotiriouHorava,WangHorava}, and explicit chronology-protection mechanisms in string-theoretic settings can also induce low-energy consequences~\cite{IRCTCstring}.

Suggestively, there already exist infrared or effective models that were introduced for other reasons yet nevertheless implement chronology protection. For instance, mimetic gravity~\cite{Chamseddine_2013}, originally developed to provide a geometric origin for cold dark matter, constrains a scalar field $\tau(x)$ to have a timelike gradient. When it is globally defined and single-valued, $\tau(x)$ supplies a time function, thereby implementing stable causality and forbidding CTCs.
Likewise, Ho\v{r}ava--Lifshitz gravity~\cite{PHorava}, which was originally introduced to make gravity perturbatively ghost-free and power-counting renormalizable, introduces a preferred foliation that excludes CTCs. Dark-sector physics has also been examined as a residual effect due to early-time foliation-induced diffeomorphism breaking~\cite{Magueijo:2024kks,Magueijo:2024zxz}. Treating the hypersurface deformation algebra, which presupposes a foliation and Hamiltonian evolution, as fundamental in canonical quantum-gravity approaches can also lead to unexplored models for modified gravity~\cite{Bojowald:2024naz}. The range of useful models that coincidentally exclude CTCs suggests that viable low-energy models may be constructed by elevating chronology protection itself to a guiding principle.

In this paper, we develop one such toy model in the context of metric--affine gravity. The starting point is the observation that in metric--affine GR the independent connection is determined only up to projective transformations, so an entire direction in the space of connections is pure gauge. We construct a chronology sector that breaks this projective invariance and constrains the corresponding non-metricity one-form to be timelike and locally closed. The resulting theory, which we call \emph{Metric--Affine Gravity Imposing Chronology}, \MAGIC{} for short, dynamically implements a locally hypersurface-orthogonal timelike connection mode using only geometric structures already present in metric--affine GR. In the globally exact sector, the selected mode defines a time function and hence realizes stable causality. On one branch, the \emph{local} dynamics is equivalent to the $w_{\mathrm{eff}}=-2$ branch of generalized unimodular gravity, providing a chronology-protecting, projective interpretation of this theory. This connection makes the propagating spectrum of \MAGIC{} calculable and separates the underlying geometric mechanism from any stabilizing ingredients that a phenomenological completion should eventually add.

The paper is organized as follows. In Sec.~\ref{sec:chrono}, we explain heuristically why a chronology-protecting quantum gravity may be expected to leave infrared modifications of GR. We motivate the implementation of chronology protection through stable causality, explain why strict non-degeneracy must accompany it, and review unimodular mimetic gravity and Ho\v{r}ava--Lifshitz gravity as related examples. In Sec.~\ref{sec:metricaffine}, we review the metric--affine Einstein--Hilbert action and its projective structure. In Secs.~\ref{sec:construction} and~\ref{sec:MAGIC} we construct \MAGIC{} and derive its field equations. In Sec.~\ref{sec:Ward} we identify the residual symmetry and fixed-background consistency condition. Then, in Secs.~\ref{sec:cosmology} and~\ref{sec:perturbations} we present the homogeneous field equations and an illustrative branch, the local generalized-unimodular representation, and the quadratic spectrum. We conclude with a summary, discussion, and brief outline of future directions in Sec.~\ref{sec:discussion}.

\subsection*{Notation and conventions}

We work in units with $\hbar = c = 8\pi G =1$ and metric signature $(-,+,+,+)$. The metric is denoted by $\mathbf{g}$ with components $g_{\mu\nu}$ and determinant $g$. Fraktur symbols denote tensor densities (e.g., $\mathfrak{A}$ and $\mathfrak{B}$), while boldface symbols denote tensors (e.g., $\mathbf{A}$ and $\mathbf{B}$). The Levi--Civita connection associated with $\mathbf{g}$ is written $\left\{\mathbf{g}\right\}$, and a general affine connection is denoted by $\mathbf{\Gamma}$. The covariant derivative associated with $\mathbf{\Gamma}$ is denoted by $\nabla_\mu$, while $\overset{(g)}{\nabla}_\mu$ denotes the metric-compatible Levi--Civita derivative.

\section{Chronology protection as an infrared constraint}\label{sec:chrono}

\subsection{Stable causality and time functions}
One sufficient condition for chronology protection is \emph{stable causality}. A spacetime $(\mathcal{M},\mathbf{g})$ is stably causal if it admits no closed causal curves and this property persists under arbitrarily small perturbations of the metric within the Lorentzian class.\footnote{For background on the causal hierarchy, see Refs.~\cite{Wald1984,LorentzianCausalityTheory,CausalHierarchy}.} Hawking showed that stable causality is equivalent to the existence of a \emph{global time function} that increases strictly along every future-directed causal curve~\cite{Hawking_1969}. A stably causal spacetime also admits a smooth \emph{temporal function} whose gradient is everywhere timelike~\cite{BernalSanchez2005,Minguzzi2010}. For brevity, we call such a temporal function a time function and denote it by $\phi$:
\begin{align}\label{DefinitionOfTimeFunction}
g^{\mu\nu}\partial_\mu \phi \partial_\nu \phi < 0
\qquad \text{everywhere}.
\end{align}
Such a function increases monotonically along every future-directed causal curve and therefore forbids CTCs, as illustrated schematically in Fig.~\ref{CTC}.
\begin{figure}[h]
\centering
\begin{tikzpicture}[scale=1]

  % Draw horizontal level curves and their labels
  \foreach \y in {-2,-1,0,1,2} {
    \draw[thin] (-3,\y) -- (3,\y);
    \node[right] at (3.2,\y) { $\phi = \y$};
  }

  % Draw background vector field (upward arrows starting from level curves)
  \foreach \x in {-2,-1.5,-1, -0.5,0,0.5,1,1.5,2} {
    \foreach \y in {-2,-1,0,1} {
      \draw[->,black] (\x,\y) -- ++(0,0.8);
    }
  }
  % Label for the background field
  \node[black] at (2.5,1.6) { $\partial_\mu \phi$};
  \draw[->,black, very thick] (2,1) -- ++(0,0.8);
  % Draw circle
  \draw[red, very thick] (0,0) circle (1.5);

  % Draw tangent vectors at angles 0, pi/2, pi, 3pi/2
  \draw[->,blue, very thick] (1.5,0) -- ++(0,0.7); % At 0 rad (right side)
  \node[ blue, very thick] at (1.8,0.8) { $v^\mu$};
  \fill (1.5,0) circle (2pt);
  \node[right, very thick] at (1.5,0) {$p_1$};

  \draw[->,blue, very thick] (0,1.5) -- ++(-0.7,0); % At pi/2 rad (top side)
  \node[ blue, very thick] at (-0.7,1.8) { $v^\mu$};
  \fill (0,1.5) circle (2pt);
  \node[very thick, above] at (0,1.5) {$p_2$};

  \draw[->,blue, very thick] (-1.5,0) -- ++(0,-0.7); % At pi rad (left side)
  \node[ blue, very thick] at (-1.8,-0.7) { $v^\mu$};
  \fill (-1.5,0) circle (2pt);
  \node[left, very thick] at (-1.5,0) {$p_3$};

  \draw[->,blue, very thick] (0,-1.5) -- ++(0.7,0); % At 3pi/2 rad (bottom side)
  \fill (0,-1.5) circle (2pt);
  \node[below, very thick] at (0,-1.5) {$p_4$};

  \node[ red, very thick] at (-1.3,-1.3) { $\gamma$};
  \node[ blue, very thick] at (0.8,-1.7) { $v^\mu$};

\end{tikzpicture}
\caption{A time function $\phi$ excludes CTCs. The red curve $\gamma$ is assumed to be a future-directed CTC in a spacetime region foliated by the level sets of $\phi$. Along any future-directed timelike curve, the contraction of its tangent with $\partial_\mu\phi$ has a fixed nonzero sign. The tangent $v^\mu$ to a closed loop cannot satisfy this condition globally, so $\gamma$ cannot be everywhere future-directed, contradicting the initial assumption.}
\label{CTC}
\end{figure}

Note that the local and global parts of the construction are separate. Local potentials may be defined on coordinate patches, but a single-valued time function on \emph{all} of spacetime is obtained when those potentials patch together globally, equivalently when the associated closed one-form is globally exact.

 \subsection{Non-degeneracy as a requirement}

The ``stability'' in stable causality is conceptually and technically restrictive. Stability confines perturbations \emph{within} the space of Lorentzian metrics. In particular, metric degeneracy cannot be allowed, since a degenerate metric is not Lorentzian and therefore lies outside the configuration space on which the causal hierarchy is defined. This is already implicit in Eq.~\eqref{DefinitionOfTimeFunction}, which assumes that $g^{\mu\nu}$ exists everywhere. Writing
\begin{align}\label{InverseMetric1}
g^{\mu\nu} = (-g)^{-1} \mathfrak{g}^{\mu\nu},
\end{align}
with
\begin{align}\label{InverseMetric2}
\mathfrak{g}^{\mu\nu} := -\frac{1}{3!} \epsilon^{\mu  \rho\sigma\lambda } \epsilon^{\nu \alpha \beta \gamma} g_{\rho \alpha} g_{\sigma \beta} g_{\lambda\gamma},
\end{align}
one sees immediately that $g^{\mu\nu}$ is well-defined only when
\begin{align}\label{strictNonDeg}
g \neq 0 \qquad \mathrm{everywhere.}
\end{align}
Thus, any implementation of chronology protection through a global time function in the sense of Eq.~\eqref{DefinitionOfTimeFunction} must also enforce strict metric non-degeneracy.

There is also a complementary geometric reason to regard non-degeneracy as essential. Metric degeneracy is closely tied to topology change, and Geroch-type results imply that topology change of spacelike slices forces either the appearance of CTCs or singularities~\cite{Geroch:TopologyInGR,Horowitz_1991,Borde1994,Borde:1999}.\footnote{Heuristically, a smooth timelike flow identifies nearby spacelike hypersurfaces by a diffeomorphism. A change in topology prevents such an identification, which can occur only if at least one integral curve fails to extend smoothly (a singularity forms) or closes on itself (forming a CTC).} If one aims to exclude CTCs while maintaining a nonsingular spacetime, metric degeneracy cannot be permitted as part of the physical configuration space.

The Geroch argument assumes the standard initial-value formulation of GR, in which spacetime is initially foliated by spacelike hypersurfaces generated by a smooth timelike flow. This presupposes non-degenerate initial data, since a degenerate initial slice does not define a Lorentzian Cauchy problem. Requiring the global existence of an inverse metric then extends this initial foliated structure throughout the entire spacetime.

\subsection{Infrared echoes of ultraviolet constraints}

If quantum gravity excludes spacetimes that are not stably causal, it is natural to expect this exclusion to reappear in the infrared as an effective constraint. The underlying logic is familiar from constrained path integrals in field theory: constraints on microscopic degrees of freedom often reappear at low energies through auxiliary fields, Lagrange multipliers, or modified effective interactions.

As an example, consider the generating functional for a scalar field $\sigma(x)$ in four spacetime dimensions,
\begin{align} \label{constrained}
Z[J]=\int \mathcal{D} \sigma \, \exp{\left(i S[\sigma] +i \int \mathrm{d}^4x\, J(x) \sigma(x)\right)}.
\end{align}
If the field configurations are restricted by a constraint $C[\sigma]=0$, this may be imposed directly in the path integral measure,
\begin{align} \label{constrainedWithDelta}
Z[J]=\int \mathcal{D} \sigma\, \delta\left(C[  \sigma]\right) \exp{\left(i S[\sigma] +i \int \mathrm{d}^4x\, J(x) \sigma(x)\right)}.
\end{align}
Using the Fourier representation of the $\delta$-functional,
\begin{align}
 \delta(C[\sigma])=\int \mathcal{D}\lambda \,
\exp{\left(i \int \mathrm{d}^4x\, \lambda(x) C[\sigma(x)]\right)},
\end{align}
one obtains the equivalent unconstrained path integral
\begin{align}
 Z[J]=\int \mathcal{D} \sigma \mathcal{D}\lambda \, \exp{\left(i S_{\mathrm{c}} [\sigma, \lambda] +i\int \mathrm{d}^4x J(x) \sigma(x)\right)},
\end{align}
where
\begin{align}
S_{\mathrm{c}}[\sigma, \lambda] := S[  \sigma] + \int \mathrm{d}^4x \lambda(x) C[\sigma (x)].
\end{align}
A constraint that was originally imposed on the space of field configurations $\sigma(x)$ can therefore be recast as an effective action involving an auxiliary field $\lambda(x)$. In the classical limit, this yields the original dynamics supplemented by a Lagrange-multiplier term enforcing the constraint. Terms of this general type appear in a variety of modified-gravity constructions, including mimetic gravity~\cite{Sebastiani_2017} and Einstein--{\ae}ther theory~\cite{AE,AE2}, and lead to nontrivial phenomenology.

We assume that a similar logic applies when chronology protection is imposed as a constraint on the underlying quantum-gravitational path integral. \Moustafa{However, we emphasize that this path integral argument serves only as motivation for our classical suppression mechanism.  A full approach would need to be able to control the measure in the Lorentzian path integral and, moreover, establish the set of ultraviolet-complete potential terms that would enforce the constraint.  This is well outside the scope of this work.  Instead, we simply ask whether terms in the low-energy effective action that disfavor closed timelike curves via stable causality possess useful phenomenology.  This is in keeping with a typical approach in quantum-gravity phenomenology, where a fundamental qualitative effect is postulated and then a low-energy theory also possessing that effect is tested. In this vein, stable causality is implemented through the joint requirements of a global time function and strict non-degeneracy, implemented as effective constraints in the infrared.}

\subsection{Existing examples of stable causality: unimodular mimetic and Ho\v{r}ava--Lifshitz gravity}

An instructive example is provided by unimodular mimetic gravity~\cite{UnimodularMimetic}. Although it was introduced for phenomenological reasons, its local equations constrain a scalar gradient to be timelike and the metric determinant to be nonzero---both ingredients needed for stable causality.

The unimodular--mimetic action is
\begin{align}
    S_{\mathrm{UM}} = \frac{1}{2} \int \mathrm{d}^4x \Big\{ & \sqrt{-g}\left[ R + \eta \Big(g^{\mu\nu} \partial_\mu \tau \partial_\nu \tau +1\right) \Big] \notag\\
    &-2\Lambda \left(\sqrt{-g} -\mathfrak{E} \right) \Big\} + S_{\mathrm{m}}.
\end{align}
Here the dynamical fields are $(\mathbf{g},\tau)$; the fields $(\eta,\Lambda)$ are Lagrange multipliers; $\mathfrak{E}$ is a fixed scalar density of the same weight as $\sqrt{-g}$; and $S_{\mathrm{m}}$ denotes the matter action. Variation with respect to $\eta$, $\Lambda$, $g^{\mu\nu}$, and $\tau$ yields
\begin{align}
    g^{\mu\nu} \partial_\mu \tau \partial_\nu \tau &=-1,\label{mimeticConstraint}\\
    \sqrt{-g} &= \mathfrak{E},\label{unimodularConstraint}\\
    G_{\mu\nu} +\Lambda g_{\mu\nu} + \eta \partial_\mu \tau \partial_\nu \tau  &=  T_{\mu\nu},\label{mimEins}\\
  \frac{1}{\sqrt{-g}} \partial_\mu \left( \sqrt{-g} \eta g^{\mu\nu} \partial_\nu \tau\right) &=0.\label{mimField}
\end{align}
Equation~\eqref{mimeticConstraint} locally identifies $\tau$ as a time function, while Eq.~\eqref{unimodularConstraint} fixes the metric determinant to a nonzero density. With the additional assumption that $\tau$ is globally defined and single-valued, these conditions imply stable causality. Equation~\eqref{mimEins} then takes the form of a modified Einstein equation in which the mimetic sector contributes an effective dust component, while $\Lambda$ plays the role of a cosmological constant.\footnote{That $\Lambda$ is constant follows from the Bianchi identity together with the assumption of matter conservation, $\nabla_\mu T^{\mu\nu}=0$.} In the limit $\Lambda=0$, one recovers the baseline mimetic construction in which $\tau$ behaves as pressureless dust and can mimic cold dark matter~\cite{Chamseddine_2013,Sebastiani_2017}.

A second example is Ho\v{r}ava--Lifshitz gravity~\cite{PHorava}. Here the original motivation is quite different. The theory is designed to be power-counting renormalizable while remaining free of ghosts or Ostrogradsky instabilities. Power-counting renormalizability can be achieved via higher-order terms in the gravitational action.\footnote{Renormalizability may also be achieved via other methods such as the asymptotic safety program~\cite{Eichhorn:2018yfc}.} Traditionally, however, these theories have been plagued by undesirable features such as ghosts, although this issue has recently been reconsidered~\cite{Donoghue:2021cza}. To retain higher-order terms while evading these pathologies, Ho\v{r}ava--Lifshitz gravity gives up exact Lorentz invariance, recovering it in the infrared as an approximate symmetry. Breaking boost invariance requires a distinction between space and time, so the natural geometry of Ho\v{r}ava--Lifshitz gravity is a spacetime equipped with a dynamical foliation. The leaves of the foliation are labeled by a dynamical scalar, the khronon $T(x)$. The theory therefore breaks active four-dimensional diffeomorphism invariance down to three-dimensional diffeomorphism invariance on each leaf and a one-dimensional $T$-reparameterization invariance that preserves the foliation; i.e., $T\rightarrow f(T)$ is a symmetry of the theory as long as $f(T)$ is monotonic.

With these symmetries, the infrared Ho\v{r}ava--Lifshitz action can be written covariantly. One first defines a unit timelike one-form, the {\ae}ther $u_\mu$, by
\begin{align}
u_\mu := \frac{-\partial_\mu T}{\sqrt{-g^{\alpha\beta} \partial_\alpha T \partial_\beta T}}.
\end{align}
 In terms of $u_\mu$, the two-derivative action, also known as the khronometric action, is given by
\begin{align}
    S_{\mathrm{Kh}} &= \frac{1}{16 \pi G_{\mathrm{Kh}}} \int \mathrm{d}^4x \sqrt{-g} \Big[R-2\Lambda - \alpha \Moustafa{\,} a_\mu a^\mu \notag \\ &~~~~~~~~~~~~- \beta  \left( \nabla_\mu u_\nu \right) \left( \nabla^\mu u^\nu \right)  - \lambda   \left( \nabla_\mu u^\mu  \right) ^2\Big],
\end{align}
where $a_\mu := u^\nu \nabla_\nu u_\mu$ is the acceleration.

Assuming that the khronon globally foliates spacetime, Ho\v{r}ava--Lifshitz gravity excludes spacetimes with CTCs because the khronon acts as a global time function. Although Lorentz symmetry is broken and particle dispersions are Lifshitz in the ultraviolet, excitations must still propagate forward in khronon time, preserving a well-defined global causal order.\footnote{See Ref.~\cite{Bhattacharyya:2015gwa} for a detailed discussion of causal structures in Ho\v{r}ava--Lifshitz gravity.}

These examples suggest that chronology protection may not be an accidental by-product. We therefore elevate it to a guiding principle for model building.

\section{Metric--affine gravity and projective structure}
\label{sec:metricaffine}

Before describing how we embed stable causality in metric--affine gravity, we review the metric--affine formulation of GR and, in particular, the projective structure of the Einstein--Hilbert action. These ingredients will play a central role in the chronology-protecting mechanism.

\subsection{Metric--affine Einstein--Hilbert action}

In metric--affine gravity, the metric $\mathbf{g}$ and the affine connection $\mathbf{\Gamma}$ are treated as independent geometric fields. The metric defines lengths, angles, and causal structure, while the connection defines parallel transport and curvature. Unlike in the purely metric formulation of GR, the relation between $\mathbf{g}$ and $\mathbf{\Gamma}$ is not imposed \emph{a priori}. Instead, it is determined dynamically by the field equations~\cite{Sotiriou_2007,metricAffine,ConnectionsMetricAffine}.

The metric--affine Einstein--Hilbert action is taken to be a functional of $\mathbf{g}$, $\mathbf{\Gamma}$, and matter fields $\Psi$,
\begin{align}\label{GR action}
    S_0 [\mathbf{g},\mathbf{\Gamma},\Psi]
    =
    \frac{1}{2}\int \mathrm{d}^4x \sqrt{-g} \,
\Big(
         R[\mathbf{g},\mathbf{\Gamma}] - 2\Lambda
      \Big)
     + S_{\mathrm{m}}[\mathbf{g},\Psi],
\end{align}
where
\begin{align}\label{eq:21}
    R[\mathbf{g},\mathbf{\Gamma}] := g^{\mu\nu}R_{\mu\nu}[\mathbf{\Gamma}],
    \qquad
    R_{\mu\nu}[\mathbf{\Gamma}] := {R^{\rho}}_{\mu\rho\nu}[\mathbf{\Gamma}],
\end{align}
and
\begin{align}\label{eq:22}
    {R^{\rho}}_{\sigma\mu\nu}[\mathbf{\Gamma}] :=
    \partial_{\mu}\Gamma^{\rho}_{\nu\sigma}
    -\partial_{\nu}\Gamma^{\rho}_{\mu\sigma}
    +\Gamma^{\rho}_{\mu\lambda}\Gamma^{\lambda}_{\nu\sigma}
    -\Gamma^{\rho}_{\nu\lambda}\Gamma^{\lambda}_{\mu\sigma}
\end{align}
are, respectively, the Ricci scalar, the Ricci tensor, and the Riemann tensor constructed from the independent connection.\footnote{We assume that the matter action does not depend explicitly on the connection, so that $S_{\mathrm{m}}=S_{\mathrm{m}}[\mathbf{g},\Psi]$. More general metric--affine models may allow direct dependence on $\mathbf{\Gamma}$~\cite{Sotiriou_2007}.}

Variation with respect to the metric while keeping the connection fixed yields
\begin{align}\label{Einstein}
    \frac{\delta S_0 }{\delta g^{\mu\nu}}
    =
    \frac{\sqrt{-g}}{2}
    \Big(
        G_{\mu\nu}[\mathbf{g},\mathbf{\Gamma}]
        + \Lambda g_{\mu\nu}
        - T_{\mu\nu}[\mathbf{g},\Psi]
    \Big),
\end{align}
where (note the symmetrization)
 \begin{align}\label{eq:24}
    G_{\mu\nu}[\mathbf{g},\mathbf{\Gamma}] :=
 R_{(\mu\nu)}[\mathbf{\Gamma}]
    - \frac{1}{2}R[\mathbf{g},\mathbf{\Gamma}]\,g_{\mu\nu}
\end{align}
is the Einstein tensor built from the metric and the independent connection, and
\begin{align}  \label{stressEnergyTensor}
  T_{\mu\nu}[\mathbf{g},\Psi] := -\frac{2}{\sqrt{-g}}
    \frac{\delta S_{\mathrm{m}}[\mathbf{g},\Psi]}{\delta g^{\mu\nu}}
\end{align}
is the energy-momentum tensor. The metric field equation is therefore
\begin{align}
    G_{\mu\nu}[\mathbf{g},\mathbf{\Gamma}] + \Lambda g_{\mu\nu}
    =
    T_{\mu\nu}[\mathbf{g},\Psi].
\end{align}
This reduces to the usual Einstein equation only if the Einstein tensor built from $\mathbf{\Gamma}$ coincides with the one computed from the Levi--Civita connection $\{\mathbf{g}\}$.

\subsection{Connection field equation and projective class}

The independent connection is determined by varying the action with respect to ${\Gamma^\rho}_{\mu\nu}$ while holding the metric fixed. One finds (see Appendix~\ref{App2} for details)
\begin{align}\label{GRGamma}
     \frac{\delta S_0 }{\delta {\Gamma^\rho}_{\mu\nu}}
     &=
     \frac{\sqrt{-g}}{2}
     \Big[-{T  ^\sigma }_{  \sigma\rho } g^{\mu\nu}
     + {T  ^\sigma }_{  \sigma\lambda } g^{  \lambda \nu}\delta^\mu  _\rho
      + {T^\mu}_{  \rho\sigma } g^{  \sigma \nu}
     \notag\\
     &-\frac{1}{\sqrt{-g}}\nabla  _\rho \!\left(\sqrt{-g}g^{\mu\nu}\right)
     +\frac{1}{\sqrt{-g}}\nabla  _\sigma \!\left(\sqrt{-g}g^{  \sigma \nu}\right)\delta^\mu  _\rho
      \Big],
\end{align}
where
\begin{align}\label{TorsionDefinition}
    {T^\rho }_{\mu\nu} := {\Gamma^\rho}_{\mu\nu} -
    {\Gamma^\rho }_{\nu\mu}
\end{align}
is the \emph{torsion tensor}. Setting Eq.~\eqref{GRGamma} to zero gives the connection field equation
\begin{align}\label{eq:37}
     0 =
     &-{T  ^\sigma }_{  \sigma\rho } g^{\mu\nu}
     + {T  ^\sigma }_{  \sigma\lambda } g^{  \lambda \nu}\delta^\mu  _\rho
      + {T^\mu}_{  \rho\sigma } g^{  \sigma \nu}
     \notag\\
     &-\frac{1}{\sqrt{-g}}\nabla  _\rho \!\left(\sqrt{-g}g^{\mu\nu}\right)
     +\frac{1}{\sqrt{-g}}\nabla  _\sigma \!\left(\sqrt{-g}g^{  \sigma \nu}\right)\delta^\mu  _\rho.
\end{align}
Its general solution is given by~\cite{Levi-Civita}
\begin{align}\label{GRPalatiniConnection}
    \Gamma  ^\rho_{\mu\nu}
    =
    \left\{g\right\}^{  \rho }_{\mu\nu}
    +
    \delta  ^\rho_\nu A_\mu,
\end{align}
where $\left\{g\right\}^{\rho}_{\mu\nu}$ is the Levi--Civita connection of the metric and $A_\mu$ is an arbitrary nondynamical one-form. Thus, the metric--affine Einstein--Hilbert action does not determine a unique affine connection, but rather an entire class of connections related by
\begin{align}\label{projInvarianceReview}
    \Gamma  ^\rho_{\mu\nu}
    \rightarrow
    \Gamma  ^\rho_{\mu\nu}
    +
    \delta  ^\rho_\nu A_\mu .
\end{align}
This is the well-known \emph{projective invariance} of metric--affine GR.

Substituting Eq.~\eqref{GRPalatiniConnection} into the curvature tensors,
\begin{align}
    {R^\rho}_{\sigma\mu\nu}[\mathbf{\Gamma}]
    &=
    {R  ^\rho }_{  \sigma \mu\nu}\!\left[\left\{\mathbf{g}\right\}\right]
    +
    2\delta  ^\rho_\sigma  \partial_{[\mu}A_{\nu]}, \label{eq:39}\\
    R_{\mu\nu}[\mathbf{\Gamma}]
    &=
    R_{\mu\nu}\!\left[\left\{\mathbf{g}\right\}\right]
    +
    2\partial_{[\mu}A_{\nu]}, \label{eq:40}\\
    R[\mathbf{g},\mathbf{\Gamma}]
    &=
    R\!\left[\mathbf{g},\left\{\mathbf{g}\right\}\right] =: R[\mathbf{g}]. \label{eq:41}
\end{align}
Since the antisymmetric contribution to the Ricci tensor drops out of the symmetric Einstein tensor, one finds
\begin{align}\label{eq:43}
    G_{\mu\nu}[\mathbf{g},\mathbf{\Gamma}]
    =
    G_{\mu\nu}\!\left[\mathbf{g},\left\{\mathbf{g}\right\}\right] =: G_{\mu\nu}[\mathbf{g}].
\end{align}
Hence, the Einstein tensor built from the independent connection coincides with the standard Levi--Civita Einstein tensor.

The same conclusion holds for freely falling trajectories: unparameterized geodesics are unchanged by projective transformations~\cite{Levi-Civita}. Metric--affine GR is therefore physically equivalent to metric GR as long as projective invariance remains exact.

This observation is central to what follows. The one-form $A_\mu$ is pure gauge and carries no independent physical content. A genuine departure from GR requires turning one of these projective directions from gauge into physics. This is the mechanism that will underlie \MAGIC{}.

\section{Constructing the chronology-protecting sector}\label{sec:construction}

\subsection{A determinant-sensitive dynamical constraint}

The two ingredients of stable causality are now clear: a global time function and strict metric non-degeneracy. In unimodular--mimetic gravity, non-degeneracy is imposed strongly through $\sqrt{-g}=\mathfrak{E}$. That choice is sufficient but \Moustafa{not necessarily minimal or the only possibility}. We instead make the infrared action sensitive to the degenerate limit $g=0$ while leaving the determinant dynamical.

A direct way to do so is to introduce a determinant-singular term
 \begin{align}\label{structureAction}
    S_{\mathrm{new}} = \frac{1}{2} \int \mathrm{d}^4x~ \left(-g\right)^{-n/2} \mathfrak{C}, \qquad n > 0,
\end{align}
where $\mathfrak{C}$ \Moustafa{is a constraint density}. Since $\mathrm{d}^4x$ is a scalar density of weight $-1$, the quantity $\left(-g\right)^{-n/2} \mathfrak{C}$ must be a scalar density of weight $+1$ in order for $S_{\mathrm{new}}$ to be a scalar.\footnote{Under a coordinate transformation $x^\mu \to x^{\mu'}$, a scalar density $\mathfrak{S}$ of weight $w$ transforms as $\mathfrak{S}' = J^{-w}\mathfrak{S}$, where $J :=\det(\partial x^{\mu'}/\partial x^\nu)$. The product of two densities of weights $w_1$ and $w_2$ has weight $w_1+w_2$. See Appendix~\ref{App1} for a brief review of tensor densities.} Since $g$ has weight $2$, the factor $(-g)^{-n/2}$ has weight $-n$, and therefore $\mathfrak{C}$ must carry weight $n+1$.

\Moustafa{As $g\to0$, if $\mathfrak C$ does not vanish, or if it vanishes less rapidly\footnote{\Moustafa{A sufficiently rapid vanishing of $\mathfrak C$ can cancel the singular factor $(-g)^{-n/2}$, resulting in an unsuppressed configuration.}} than $(-g)^{n/2}$, the determinant-weighted integrand $\left(-g\right)^{-n/2} \mathfrak{C}$ becomes singular. When its integrated Euclidean contribution tends to $+\infty$, the corresponding configurations are exponentially suppressed: $e^{-S} \to 0$.}

If $\mathfrak{C}$ is further constructed so that its vanishing enforces $v^\mu \nabla_\mu \sqrt{-g} >0$, with $v^\mu$ being directed toward a degeneracy point, then $\sqrt{-g}$ is dynamically pushed away from degeneracy along any integral curve of $v^\mu$. This mechanism necessarily lies outside purely metric GR, because it requires a connection for which $\nabla_\mu \sqrt{-g}$ does not vanish identically. We are therefore naturally led to metric--affine gravity.

The second ingredient is the existence of a time function. Ideally, both requirements would emerge from the same new term, so that non-degeneracy and chronology protection are linked at the level of the action. This suggests introducing a Lagrange multiplier that constrains the norm of a geometrically natural one-form constructed from the affine structure. We then construct $\mathfrak{C}$ such that the condition $\mathfrak{C} = 0$ contributes to the field equations a constraint equivalent to Eq.~\eqref{DefinitionOfTimeFunction}. 

\Moustafa{As an attempt at constructing $\mathfrak{C}$, we choose}
\begin{align}\label{newTerm}
    &S_{\mathrm{new}} := \notag\\
    & \frac{1}{2} \int\mathrm{d}^4x \Moustafa{\,} \lambda \left(-g\right)^{-3/2} \left[ \mathfrak{g}^{\mu\nu} \left(\nabla_\mu \sqrt{-g}\right) \left(\nabla_\nu \sqrt{-g}\right) + \mathfrak{m}^2 \right],
\end{align}
where $\lambda(x)$ is a Lagrange multiplier, $\mathfrak{g}^{\mu\nu}$ is the inverse metric density defined in Eqs.~\eqref{InverseMetric1}--\eqref{InverseMetric2}, and $\mathfrak{m}(x)$ is a scalar density of the same weight as $g$, namely, weight~$2$. \Moustafa{Compared to Eq.~\eqref{structureAction},
\begin{align}\label{newC}
    \mathfrak{C}_{\mathrm{new}} = \lambda \left[ \mathfrak{g}^{\mu\nu} \left(\nabla_\mu \sqrt{-g}\right) \left(\nabla_\nu \sqrt{-g}\right) + \mathfrak{m}^2 \right].
\end{align}
As we will show below, this selects a class of regular non-degenerate branches under an additional orientation condition.\footnote{\Moustafa{Since this is a minimal construction, it allows exceptional approaches to degeneracy that we discuss separately in Sec.~\ref{subsec:exceptions}. For example, the construction in Eq.~\eqref{newC} does not guarantee that $\mathfrak{C}$ remains non-negative.}}}

Here, we take $\sqrt{\mathfrak{m}}$ to be the component density of a smooth, positive, and \emph{fixed} geometric volume form $\boldsymbol{\omega}_{\mathfrak{m}}$,
\begin{align}\label{fixed volume form}
    \boldsymbol{\omega}_{\mathfrak{m}}
    := \sqrt{\mathfrak{m}}\,\mathrm{d}^4x.
\end{align}
\Moustafa{In terms of $\boldsymbol{\omega}_{\mathfrak{m}}$, the metric volume form can be written geometrically as
\begin{align}\label{volume_relation}
\sqrt{-g}\,\mathrm d^4x= \frac{\boldsymbol\omega_{\mathfrak m}}{\sqrt{m}},
\end{align}}
where 
\begin{align}  \label{ratio}
  \sqrt{m} := \frac{\sqrt{\mathfrak{m}}}{\sqrt{-g}}.
\end{align}
\Moustafa{Metric-volume collapse, or metric degeneracy, is then invariantly characterized by $\sqrt{m} \to \infty$.}

The consistency condition associated with the existence of this fixed geometric object is derived in Sec.~\ref{sec:Ward}. For now, we point out that the component of $\boldsymbol{\omega}_{\mathfrak{m}}$---that is $\sqrt{\mathfrak{m}}$---transforms under \emph{passive} coordinate changes. The mere existence of $\boldsymbol{\omega}_{\mathfrak{m}}$, however, reduces the \emph{active} gauge symmetry and hence changes the associated Ward identity. Moreover, different coordinate profiles of $\mathfrak{m}(x)$ may represent the \emph{same} fixed volume form. Indeed, one may choose local volume-adapted coordinates in which $\boldsymbol{\omega}_{\mathfrak{m}}=\mathrm{d}^4y$, effectively setting $\mathfrak{m}(y)=1$. The local component profile of a given form therefore has no invariant content by itself\Moustafa{---only $\boldsymbol{\omega}_{\mathfrak{m}}$ is geometrically meaningful.}

Two additional structural features of Eq.~\eqref{newTerm} are important. First, unlike the inverse metric $g^{\mu\nu}$ which becomes singular at $g=0$, the density $\mathfrak{g}^{\mu\nu}$ is finite. This is particularly clear from Eq.~\eqref{InverseMetric2}. Second, although the suppression of degeneracy is most directly relevant near $g=0$, the Lagrange multiplier constraint acts throughout the whole domain---both degenerate and non-degenerate.

The same multiplier term therefore links sensitivity to metric degeneracy with the causal structure of non-degenerate configurations. The class of branches dynamically repelled from $g=0$ is characterized in the next subsection.

 \subsection{Regular non-degenerate branches}

\Moustafa{Variation with respect to $\lambda$ requires the square bracket in Eq.~\eqref{newTerm} to vanish:}
\begin{align}\label{eq:density_constraint}
    \mathfrak{g}^{\mu\nu} \left(\nabla_\mu \sqrt{-g}\right) \left(\nabla_\nu \sqrt{-g}\right) +\mathfrak{m}^2 = 0.
 \end{align}
If $\nabla_\mu\sqrt{-g}\neq0$, which must be the case for $\mathfrak{m} \neq 0$, one may formally write the solution for $\nabla_\mu \sqrt{-g}$ as
\begin{align}\label{5.4}
    \nabla_\mu \sqrt{-g} = \sqrt{\mathfrak{m}}~\tilde{Q}_\mu,
\end{align}
for some nonvanishing one-form $\tilde{Q}_\mu$. Equation~\eqref{5.4} parametrizes any solution of Eq.~\eqref{eq:density_constraint} as $\sqrt{\mathfrak{m}} \,\tilde{Q}_\mu$. The particular form of $\tilde{Q}_\mu$ is not important for the degeneracy argument; the only requirement is that $\tilde{Q}_\mu$ remains nonzero and properly oriented as $\sqrt{-g} \to 0$. The factor of $\sqrt{\mathfrak{m}}$ ensures that the right-hand side transforms as a density of weight 1.

Let $\gamma$ be a curve parameterized by $\sigma \in [0,1]$ with tangent vector $v^\mu := \left(\mathrm{d}x^\mu / \mathrm{d}\sigma\right)|_\gamma$. Suppose that along $\gamma$ the determinant tends toward zero from above in the direction of increasing $\sigma$, so that $\sqrt{-g(1)}=0$ would correspond to a would-be degenerate endpoint. Contracting Eq.~\eqref{5.4} with $v^\mu$ yields the covariant directional derivative
\begin{align}\label{5.5}
    v^\mu \nabla_\mu \sqrt{-g} = \sqrt{\mathfrak{m}} \left(v^\mu \tilde{Q}_\mu\right).
\end{align}
If $v^\mu \tilde{Q}_\mu > 0$, the geometry is pushed away from $g=0$ along that curve.

Since the covariant derivative of a scalar density $\mathfrak{S}$ of weight $w$ is (see Appendix~\ref{App1})
\begin{align}\label{A5}
    \nabla_\mu \mathfrak{S} := \partial_\mu \mathfrak{S} -w \Gamma^\nu_{\mu\nu} \mathfrak{S},
\end{align}
Eq.~\eqref{5.5} yields the inhomogeneous differential equation
\begin{align}
    \frac{\mathrm{d} \sqrt{-g}}{\mathrm{d}\sigma} - v^\mu \Gamma^\nu_{\mu\nu} \sqrt{-g} = \sqrt{\mathfrak{m}} \left(v^\mu \tilde{Q}_\mu\right),
\end{align}
whose general solution is
\begin{align}\label{eq:5.7}
    &\sqrt{-g(\sigma)} = \exp{\left(\int_0^\sigma v^\mu \Gamma^\nu_{\mu\nu} \mathrm{d}\sigma'\right)} \Bigg[ \sqrt{-g(0)} \notag \\
    & +\int_0^\sigma \mathrm{d} \sigma' \left(\sqrt{\mathfrak{m}} v^\mu \tilde{Q}_\mu\right) \exp{\left(-\int_0^{\sigma'} v^\mu \Gamma^\nu_{\mu\nu} \mathrm{d}{\sigma''}\right)} \Bigg].
\end{align}
For any regular but otherwise arbitrary connection $\mathbf{\Gamma}$ \Moustafa{with finite $\int_0^1 v^\mu\Gamma^\nu_{\mu\nu}\,\mathrm d\sigma$}, if $\sqrt{-g(0)} > 0$ and $v^\mu \tilde{Q}_\mu>0$ along every direction leading toward a would-be degenerate point, Eq.~\eqref{eq:5.7} ensures that $\sqrt{-g(\sigma)}>0$.

This identifies a dynamically protected class of regular branches, namely those for which the orientation condition $v^\mu\tilde Q_\mu>0$ holds along every approach to metric degeneracy. One should be careful, however: although this is a sufficient orientation condition, it is not an independent equation imposed on every solution.

\subsection{From non-metricity to a timelike one-form}

Variation of Eq.~\eqref{newTerm} with respect to $\lambda$ yields Eq.~\eqref{eq:density_constraint}, which already resembles the defining equation for a local time function, Eq.~\eqref{DefinitionOfTimeFunction}. However, it is not yet in a coordinate-invariant form. The problem is that $\sqrt{-g}$ is a scalar \emph{density} rather than a proper scalar, so its level sets do not by themselves define a diffeomorphism-invariant foliation.

This distinction can be seen by comparing the foliations defined by
\begin{align}\label{foliation1}
    \phi(x) = C,
\end{align}
and
\begin{align}\label{foliation2}
    \sqrt{-g(x)} = C.
\end{align}
Under a coordinate transformation $x^\mu \rightarrow x^{\mu'}$, the first foliation remains invariant since both sides of Eq.~\eqref{foliation1} are scalars. The second foliation, however, transforms as
\begin{align}\label{transformedFoliation2}
    J^{-1}(x) \sqrt{-g(x)} = C,
\end{align}
where $J(x)$ is the Jacobian determinant of the transformation. Only when $J(x)=1$ does Eq.~\eqref{transformedFoliation2} coincide with Eq.~\eqref{foliation2}. A foliation defined by a scalar density is therefore not invariant under general coordinate transformations, but only under the subgroup of volume-preserving coordinate transformations.

To ensure full coordinate invariance, a foliation must be defined by a proper scalar field, which we construct \textit{solely from gravitational structure}. On a non-degenerate branch, we may divide Eq.~\eqref{eq:density_constraint} by $g^2$ and use Eq.~\eqref{InverseMetric1} to obtain
\begin{align}\label{5.19}
    g^{\mu\nu}  \left( \frac{\nabla_\mu \sqrt{-g}}{\sqrt{-g}}\right) \left( \frac{\nabla_\nu \sqrt{-g}}{\sqrt{-g}}\right) = - m^2.
\end{align}
The natural geometric quantity appearing in Eq.~\eqref{5.19} is the \emph{non-metricity}. The \emph{non-metricity tensor} is defined as
\begin{align}  \label{nonMetrDef}
Q_{\rho\mu\nu} := \nabla_\rho g_{\mu\nu},
\end{align}
and the non-metricity one-form is its trace
\begin{align}\label{nonMetrVector}
Q_\rho := g^{\mu\nu}Q_{\rho\mu\nu} = g^{\mu\nu}\nabla  _\rho  g_{\mu\nu} =
2\frac{\nabla_\rho \sqrt{-g}}{\sqrt{-g}}.
\end{align}
Equation~\eqref{5.19} is therefore equivalent to
\begin{align}\label{normNonMetr}
g^{\mu\nu}Q_\mu Q_\nu = -4m^2.
\end{align}
Thus, the new term forces the non-metricity one-form to be timelike.

\subsection{Twist-free non-metricity, a local time function, and global exactness}

A timelike one-form is not yet enough to define a time function. We also require hypersurface orthogonality, or equivalently vanishing twist. We therefore impose
\begin{align}\label{twistFree}
    \partial_{[\mu} Q_{\nu]} =0.
\end{align}
When this condition holds, one locally has
\begin{align}\label{WeylVector}
Q_\mu = 2 \partial_\mu \phi
\end{align}
for some scalar field $\phi$. The norm constraint, Eq.~\eqref{normNonMetr}, then becomes
\begin{align}\label{7.13}
g^{\mu\nu}\partial_\mu\phi\,\partial_\nu\phi=-m^2.
\end{align}
Thus, locally, $\phi$ has a timelike gradient.

A global time function requires the additional condition that the closed one-form $Q_\mu$ be globally exact, so that
a single-valued scalar $\phi$ exists on all of $\mathcal{M}$. If its level sets define a global foliation under suitable completeness assumptions, the spacetime may further admit a product
decomposition $\mathcal{M} \simeq \mathbb{R} \times \Sigma$. Because we consider only
the infrared limit of a postulated quantum-gravity effect,
we do not address how this global requirement should be
extended to possible topology changes in quantum gravity. Throughout the
remainder of the paper, writing a single scalar $\phi$ assumes
global exactness. Without this input, the action enforces
only a local timelike foliation.

Thus, the chronology sector extracts a timelike closed one-form directly from the non-metricity of the independent connection and, in the globally exact sector, promotes it to a global time function.

\subsection{Why the Levi--Civita connection is insufficient}\label{The Need for a New Connection}

The construction above is impossible with the Levi--Civita connection. Since
\begin{align}
\overset{(g)}{\nabla}_\rho g_{\mu\nu} :=0,
\end{align}
one consequently has
\begin{align}
\overset{(g)}{\nabla}_\mu\sqrt{-g}=0,
\end{align}
so Eq.~\eqref{eq:density_constraint} cannot be satisfied when $\mathfrak{m}\neq0$. The chronology-protecting sector therefore requires a genuinely non-metric affine connection.

The twist-free condition on $Q_\mu$ places a direct restriction on the allowed connection. Using Eq.~\eqref{A5}, the non-metricity one-form in Eq.~\eqref{nonMetrVector} can be expressed as
\begin{align}
  Q_\mu = 2\left( \frac{\partial_\mu \sqrt{-g}}{\sqrt{-g}} - \Gamma  ^\rho_{\mu  \rho }\right).
\end{align}
The condition \(\partial_{[\mu}Q_{\nu]}=0\) then implies
\begin{align}\label{gradient constraint}
\partial_{[\mu}\Gamma  ^\rho_{\nu]  \rho }=0.
\end{align}
Equivalently, the trace of the affine connection differs from the Levi--Civita trace by a gradient,
\begin{align}\label{differenceBetweenConnections}
  \partial_\mu \phi = \left\{g\right\}  ^\rho_{\mu  \rho } - \Gamma  ^\rho_{\mu  \rho }.
\end{align}
This departure from the Levi--Civita connection is the structure required to generate a local time function.

A useful equivalent formulation is obtained in terms of the \emph{homothetic curvature},
\begin{align}\label{homothetic}
    R'_{\mu\nu}[\mathbf{\Gamma}] :={R^\rho}_{\rho\mu\nu}[\mathbf{\Gamma}] = 2\partial_{[\mu}\Gamma^\rho_{\nu]\rho}.
\end{align}
We therefore obtain the chain of equivalent statements
\begin{align}\label{equivalentConditions}
    \partial_{[\mu} Q_{\nu]} =0 \quad \Longleftrightarrow \quad \partial_{[\mu}\Gamma  ^\rho_{\nu]  \rho } =0 \quad \Longleftrightarrow \quad R'_{\mu\nu}[\mathbf{\Gamma}] =0.
\end{align}

\subsection{Breaking projective invariance}

The crucial structural point is now clear. In metric--affine GR, the one-form appearing in the projective transformation of the connection is pure gauge. The chronology-protecting sector we introduce breaks this projective invariance and ties the physically relevant projective mode to the gradient of a scalar time function.

The would-be gauge direction thereby becomes constrained geometric structure imposing chronology protection rather than requiring the introduction of an independent matter field. In this sense the construction changes the \emph{status} of a projective direction already present in metric--affine GR without enlarging the list of matter fields.

\Moustafa{\subsection{Degenerate-limit exceptions and effective actions} \label{subsec:exceptions}
We have motivated, although not derived, our action from the idea that there may be an echo of chronology protection in the infrared limit.  However, precisely because we are not dealing with an explicit derivation from a complete quantum gravity theory, we must be careful about how far we can push the claim of metric-degeneracy suppression. Our proposed term is not necessarily the \textit{only} term that can be involved in degeneracy suppression at all energy scales.  Other terms may come into play at high energies or when geometric invariants grow large. Moreover, the treatment of degenerate metrics in a gravitational path integral depends on the integration domain, measure, and contour~\cite{BanihashemiJacobson2025Lapse,Thiemann2024Nondegenerate}.  In particular, because $\lambda$ enters linearly, a Euclidean suppression interpretation requires a prescription for the $\lambda$-integration contour that renders the relevant contribution real and positive. Since we do not specify such a prescription here, we regard the suppression discussion below as motivation. The following discussion illustrates where other terms may play a role.

The use of scalar densities may occasionally blur the coordinate-independent content of the construction. We therefore use Eqs.~\eqref{volume_relation},~\eqref{ratio}, and~\eqref{nonMetrVector} to rewrite Eq.~\eqref{newTerm} in terms of proper scalars:
\begin{align} \label{eq:snew}
S_{\mathrm{new}}=\frac12\int\boldsymbol\omega_{\mathfrak m}\,
\lambda m^{3/2}\left(1+\frac{Q_\mu Q^\mu}{4m^2}\right).
\end{align}
Metric-volume collapse relative to the fixed volume form therefore corresponds to $m\to\infty$, which is a meaningful geometric statement. When $Q_\mu Q^\mu$ remains finite and $\lambda$ approaches a positive nonzero limit, the integrand grows as $\lambda m^{3/2}$. This gives exponential suppression when the chosen continuation and contour yield a positive divergent integrated Euclidean action.\footnote{\Moustafa{For $\lambda<0$, the same divergent integrated contribution would have the opposite sign.}}

There is a scenario, however, where the scalar in parentheses may vanish identically even as $m$ diverges; this is the case when $Q^\mu Q_\mu$ diverges exactly as $-4m^2$, so that $Q^\mu Q_\mu / 4m^2 =-1$. The multiplier term then vanishes identically and supplies no bulk-action penalty. As an example, this occurs for the following family of metrics\footnote{This example was supplied by an anonymous referee.}
\begin{align}
    ds_\epsilon^2=-(t+\epsilon)^2 \mathrm{d}t^2 +\mathrm{d}\mathbf{x}^2,
    \quad 0 \leq t \leq 1,\quad \epsilon>0,
\end{align}
with $\mathfrak m=1$ and $\Gamma^\rho{}_{\mu\nu}=0$. For this metric--affine configuration,
\begin{align}
    m=\frac{1}{(t+\epsilon)^2},\qquad
    Q_\mu\mathrm dx^\mu=\frac{2\,\mathrm dt}{t+\epsilon},\qquad
    Q_\mu Q^\mu=-4m^2.
\end{align}
The parentheses in Eq.~\eqref{eq:snew} consequently vanish for every $\epsilon>0$, even though both $m$ and $Q_\mu Q^\mu$ diverge at $t=0$ as $\epsilon\to0$. With $\xi^{[\mu\nu]}=0$, $S_{\mathrm m}=0$, and $\Lambda=0$, the remaining bulk terms of the \MAGIC{} action vanish as well because $R[\mathbf\Gamma]=0$. This off-shell family therefore has no divergent bulk-action penalty in the theory as written.

Indeed, this example lies outside the class protected by Eq.~\eqref{eq:5.7} for which we established classical non-degeneracy. Equation~\eqref{5.4} gives $\tilde Q=\mathrm dt$, so a curve directed toward $t=0$ has $v^\mu\tilde Q_\mu=v^t<0$, whereas our sufficient orientation condition requires $v^\mu\tilde Q_\mu>0$. It is therefore a counterexample to an \emph{unconditional} action-level barrier, but not to the conditional classical result for regular, appropriately oriented branches.

Still, this does illuminate an essential point: under the Euclidean conditions stated above, the \MAGIC{} action can suppress \emph{some}, but not \emph{all}, approaches to degeneracy---the tuned family above evades the barrier. The absence of a bulk-action penalty for this family---with divergent but \textit{tuned} $Q_\mu Q^\mu$ and $m^2$---is sensitive to perturbations of the action.

A generic metric--affine effective action contains independent contractions and higher powers of non-metricity~\cite{Burgess2004EFT,Iosifidis2022QuadraticMAG}; for instance, one can have nonlinear functions of a non-metricity scalar~\cite{BeltranJimenez2018Coincident,BeltranJimenez2020fQ}. A positive divergent integrated Euclidean contribution from such terms would suppress the tuned family.\footnote{A symmetry could restrict such operators or enforce further cancellations; none is assumed here.}

For concreteness, on the family above, a quartic-non-metricity term has, per unit coordinate spatial volume,
\begin{align}
\int_0^1\mathrm dt\,\sqrt{-g}\,(Q_\mu Q^\mu)^2
=16\int_0^1\frac{\mathrm dt}{(t+\epsilon)^7}
\sim\frac{8}{3\epsilon^6}.
\end{align}
An additional term of this type detects the tuned approach to degeneracy and, with a positive real Euclidean coefficient, suppresses it even while $Q_\mu Q^\mu=-4m^2$ holds exactly.

Such additional polynomial terms would be near-vanishing for small $Q_\mu Q^\mu$ and not typically affect the equations of motion in the near-GR limit, which is where we focus in this work. Hence, we leave to future work the (admittedly interesting) question of perturbations of the \MAGIC{} action that would additionally forbid tuned divergences. For now, we accept such configurations as exceptions to unconditional bulk-action suppression in our minimal \MAGIC{} model.}

\section{Metric--Affine Gravity Imposing Chronology (\MAGIC{})}\label{sec:MAGIC}

\subsection{Action}

We now present the full \MAGIC{} action. Two terms are added to the Einstein--Hilbert action. The first, introduced in Eq.~\eqref{newTerm}, is degeneracy-sensitive, imposes the timelike norm constraint on the non-metricity one-form, and selects the regular branches characterized above. The second imposes the necessary complementary condition that the homothetic curvature vanish, which makes the non-metricity one-form twist-free. Thus,
\begin{align}\label{Action}
    &S_{\mathrm{\MAGIC{}}} = \frac{1}{2}\int \mathrm{d}^4x \sqrt{-g} \,\Big(R[\mathbf{g},\mathbf{\Gamma}] -2\Lambda\Big) + S_{\mathrm{m}}[\mathbf{g},\Psi;\mathfrak{m}]\notag\\
    &+ \frac{1}{2}\int \mathrm{d}^4x~ \lambda \left(-g\right)^{-3/2} \left[ \mathfrak{g}^{\mu\nu} \left(\nabla_\mu \sqrt{-g}\right) \left(\nabla_\nu \sqrt{-g}\right) + \mathfrak{m}^2 \right]\notag\\
    &+ \frac{1}{2} \int \mathrm{d}^4x \sqrt{-g} \, \xi^{ [ \mu\nu ] } R'_{\mu\nu}.
\end{align} 
The full action is a functional of $(\mathbf{g}, \mathbf{\Gamma},\Psi,\lambda,\boldsymbol{\xi};\mathfrak{m})$, where $\lambda$ and $\boldsymbol{\xi}$ are Lagrange multipliers. The semicolon separates the fixed volume density from the dynamical fields. Note that because $R'_{\mu\nu}$ is antisymmetric, only $\xi^{[\mu\nu]}$ is operative; the symmetric part $\xi^{(\mu\nu)}$ is absent and is not a field of the theory.

Another subtle point is that, when discussing metric--affine GR in Sec.~\ref{sec:metricaffine}, we assumed that the matter action depends only on $\Psi$ and $\mathbf{g}$. Here, however, in constructing \MAGIC{}, we further allow dependence on $\mathfrak{m}$. In this way, both the matter and gravitational sectors share the same residual diffeomorphism symmetry. We discuss this point in more detail in Sec.~\ref{sec:Ward}.

Variation with respect to $\lambda$ imposes Eq.~\eqref{eq:density_constraint}, while variation with respect to $\xi^{[\mu\nu]}$ imposes Eq.~\eqref{gradient constraint} or equivalently $R'_{\mu\nu}=0$. 

The $\xi^{[\mu\nu]}$ sector has the reducible gauge symmetry
\begin{align}\label{xi-gauge}
  \xi^{[\mu\nu]}
    \longrightarrow
    \xi^{[\mu\nu]}
    +\frac{1}{\sqrt{-g}} \partial_\rho\left(\sqrt{-g} \, \Sigma^{[\rho\mu\nu]}\right),
\end{align}
with gauge-for-gauge transformation
\begin{align}\label{xi-reducibility}
  \Sigma^{[\rho\mu\nu]}
    \longrightarrow
    \Sigma^{[\rho\mu\nu]}
    +\frac{1}{\sqrt{-g}} \partial_\sigma \left(\sqrt{-g}\, \Upsilon^{[\sigma\rho\mu\nu]}\right).
\end{align}
This reducible symmetry is important for the degree-of-freedom count.

Finally, the linear homothetic-curvature coupling distinguishes \MAGIC{} from projective-breaking theories with nonlinear curvature terms~\cite{BeltranJimenezDelhom2019,BeltranJimenezDelhom2020}. Here the $\xi^{[\mu\nu]}$ equation sets $R'_{\mu\nu}=0$, so the higher-curvature kinetic structure studied in those works is absent from the classical action.

\subsection{Connection, torsion, and non-metricity}

Varying the action with respect to ${\Gamma^\rho}_{\mu\nu}$ yields 
\begin{align}\label{eq:7.5}
      \frac{\delta S_0 }{\delta {\Gamma^\rho}_{\mu\nu}}
     -  \delta^\nu_\rho
     \left[\lambda g^{\mu\sigma}\nabla_\sigma \sqrt{-g} + \partial_\sigma\!\left(\sqrt{-g}\xi^{[\sigma\mu]}\right)
    \right]=0.
\end{align}
Contracting Eq.~\eqref{eq:7.5} with $\delta^\rho_\nu$ gives
\begin{align}\label{7.4}
    \delta^\rho_\nu
    \frac{\delta S_0 }{\delta {\Gamma^\rho}_{\mu\nu}}
     = 4 \left[
          \lambda g^{\mu\sigma}\nabla_\sigma \sqrt{-g}
        +\partial_\sigma\!\left(\sqrt{-g}\xi^{[\sigma\mu]}\right)
    \right].
\end{align}
From Eq.~\eqref{GRGamma}, one finds the off-shell identity
\begin{align}\label{eq:65}
    \delta^\rho_\nu \frac{\delta S_0 }{\delta {\Gamma^\rho}_{\mu\nu}} \equiv 0.
\end{align}
Substituting Eq.~\eqref{eq:65} back into Eqs.~\eqref{7.4} and~\eqref{eq:7.5} gives
\begin{subequations}
    \begin{align}
         0 &= \frac{\delta S_0 }{\delta {\Gamma^\rho}_{\mu\nu}}, \label{eq:66a} \\
        0 &= \lambda g^{\mu\nu}\nabla_\nu \sqrt{-g}
        +\partial_\nu\!\left(\sqrt{-g}\xi^{[\nu\mu]}\right). \label{eq:71b}
    \end{align}
\end{subequations}

Equation~\eqref{eq:71b} can be written as
\begin{align}\label{Maxwell_eq}
\partial_\nu \mathfrak{F}^{\mu\nu} = \mathfrak{J}^\mu,
\end{align}
with a Maxwell-like tensor density
\begin{align}
  \mathfrak{F}^{\nu\mu} := \sqrt{-g}\,\xi^{[\nu\mu]} = - \mathfrak{F}^{\mu\nu},
\end{align}
and a vector-density source, which we call the \emph{projective current},
\begin{align}\label{source}
  \mathfrak{J}^\mu := \lambda g^{\mu\nu} \nabla_\nu \sqrt{-g}
    = \frac{1}{2}\lambda \sqrt{-g} \, Q^\mu.
\end{align}
Applying $\partial_\mu$ to both sides of Eq.~\eqref{Maxwell_eq} and using the antisymmetry of $\mathfrak{F}^{\nu\mu}$, we obtain the consistency condition
\begin{align}\label{connection-integrability-Q}
\partial_\mu \mathfrak{J}^\mu =0.
\end{align}
Thus, Eq.~\eqref{eq:71b} restricts the projective current to be conserved.\footnote{\Moustafa{Equation~\eqref{connection-integrability-Q} is \emph{locally} sufficient to reconstruct $\mathfrak F^{\mu\nu}$ on a contractible patch; \emph{globally}, however, the associated closed current three-form must also be exact, with boundary data admitting the required flux. In particular, a globally smooth multiplier on a compact spatial slice $\Sigma$ without boundary requires $\int_\Sigma\mathrm d^3x\,\mathfrak J^\tau=0$. This restriction is distinct from global exactness of $Q_\mu$.}} 

Once the constraint Eq.~\eqref{connection-integrability-Q} is imposed, the remaining connection equation, Eq.~\eqref{eq:66a}, is the Palatini equation and retains the solution in Eq.~\eqref{GRPalatiniConnection}. Taking Eq.~\eqref{differenceBetweenConnections} into account, we obtain the \MAGIC{} connection
 \begin{align}\label{NewConnection}
    \Gamma^{  \rho }_{\mu\nu} = \left\{g\right\}^{  \rho }_{\mu\nu} - \frac{1}{4} \delta  ^\rho_\nu \partial_\mu \phi.
\end{align}
It follows from Eq.~\eqref{TorsionDefinition} that the torsion tensor and its trace are
 \begin{align}
    {T^\rho }_{\mu\nu}  & = \frac{1}{4} \left(\delta^\rho_\mu \partial_\nu \phi -\delta^\rho_\nu \partial_\mu \phi\right),  \\
  T_\mu & := {T^\nu }_{\mu  \nu } =-{T  ^\nu }_{  \nu \mu} = -\frac{3}{4} \partial_\mu \phi. \label{TorsionTrace}
 \end{align}

For the \MAGIC{} connection in Eq.~\eqref{NewConnection}, the covariant derivative of an $(n,l)$-tensor density $\boldsymbol{\mathfrak{T}}$ of weight $w$ is related to the Levi--Civita derivative by (see Eq.~\eqref{A7} in Appendix~\ref{App1}) \begin{align}\label{RelationBetweenNewAndOldCovariantDerivative}
    \nabla  _\rho  {\mathfrak{T}^{\mu_1 \dots \mu_n}}_{\nu_1 \dots \nu_l} &= \overset{(g)}{\nabla}  _\rho  {\mathfrak{T}^{\mu_1 \dots \mu_n}}_{\nu_1 \dots \nu_l} \notag \\
    &- \frac{1}{4}\left(n-l-4w\right) \left(\partial  _\rho  \phi\right) {\mathfrak{T}^{\mu_1 \dots \mu_n}}_{\nu_1 \dots \nu_l}.
\end{align}
Together, Eqs.~\eqref{nonMetrDef} and~\eqref{RelationBetweenNewAndOldCovariantDerivative} then give
 \begin{align}
        Q_{  \rho \mu\nu} = \frac{1}{2} \left(\partial  _\rho  \phi\right) g_{\mu\nu},
\end{align}
and thus both Eqs.~\eqref{WeylVector} and~\eqref{7.13} remain satisfied.

The connection selected by \MAGIC{} therefore carries a timelike, twist-free non-metricity and hence the local chronology-protecting structure.

\subsection{Modified gravitational field equations}
Varying with respect to $g^{\mu\nu}$ and setting the result to zero yields (see Appendix~\ref{App4} for details)
\begin{align}\label{NewEinstein}
     G_{\mu\nu} + \Lambda g_{\mu\nu} +\mathcal{C}_{\mu\nu} =T_{\mu\nu},
\end{align}
where $G_{\mu\nu}$ is the ordinary Einstein tensor built from the Levi--Civita connection and $\mathcal{C}_{\mu\nu}$ is the chronology-sector correction, interpreted here as a possible infrared imprint of quantum gravity,
 \begin{align}  \label{Delta-QG-reduced}
  \mathcal{C}_{\mu\nu} := \lambda \partial_\mu \phi \partial_\nu \phi + 2 \lambda m^2 g_{\mu\nu},
\end{align}
and the scalar field $\phi$ is restricted to satisfy
\begin{align}\label{phiConservation}
  \frac{1}{\sqrt{-g}} \partial_\mu \left( \lambda \sqrt{-g} g^{  \mu\nu } \partial  _\nu  \phi\right) =0.
 \end{align}
Equation~\eqref{phiConservation} follows directly from substituting Eqs.~\eqref{WeylVector} and~\eqref{source} into Eq.~\eqref{connection-integrability-Q}.

 It is useful to introduce the unit timelike one-form
\begin{align}  \label{unitTimelike}
  u_\mu := -\frac{\partial_\mu \phi}{m}, \qquad u^\mu u_\mu =-1.
\end{align}
In later sections, we will examine the cosmological implications in an FLRW scenario. To do so, it is convenient to move the correction to the right-hand side and define the \emph{formal} effective tensor
\begin{align}
  T^{\mathrm{(\MAGIC{})}}_{\mu\nu}
    :=-\mathcal{C}_{\mu\nu},
\end{align}
which may be written in perfect-fluid form,
\begin{align}
    T^{\mathrm{(\MAGIC{})}}_{\mu\nu} = \left(\varepsilon_{\mathrm{eff}} + p_{\mathrm{eff}}\right) u_\mu u_\nu + p_{\mathrm{eff}} g_{\mu\nu}, \label{PerfectFluid}
\end{align}
with
\begin{align}
     \varepsilon_{\mathrm{eff}} := \lambda m^2, \qquad p_{\mathrm{eff}}:= -2\lambda m^2.\label{density_pressure}
\end{align}
Thus, the nontrivial homogeneous branch has the formal equation-of-state parameter
\begin{align}
  w_{\mathrm{eff}}:=
    \frac{p_{\mathrm{eff}}}{\varepsilon_{\mathrm{eff}}}=-2.
\end{align}
This value is \emph{fixed} on the nontrivial local branch. For $\lambda>0$, the formal effective source has positive energy density but violates the null energy condition; for $\lambda<0$, its energy density is negative. The branch $\lambda=0$ carries no such effective source.

\subsection{Regularity of the preferred congruence}
\label{sec:foliation-regularity}

The unit timelike one-form introduced in Eq.~\eqref{unitTimelike} defines the preferred congruence associated with the foliation. Its kinematics expose a nonlinear regularity question that is not settled by the local norm and closure constraints. Defining the spatial projector
\begin{align}
  h_\mu{}^\nu := \delta_\mu{}^\nu+u_\mu u^\nu,
\end{align}
the Levi--Civita acceleration of the congruence is
\begin{align}\label{aether-acceleration}
  a_\mu
    := u^\nu\overset{(g)}{\nabla}_\nu u_\mu
    =-h_\mu{}^\nu\partial_\nu\ln m.
\end{align}
The congruence is therefore geodesic whenever the spatially projected gradient of $m$ vanishes. Since the twist vanishes by construction, the Raychaudhuri equation reduces to
\begin{align}\label{raychaudhuri-MAGIC}
  u^\nu\overset{(g)}{\nabla}_\nu\theta
    =-\frac13\theta^2
    -\sigma_{\mu\nu}\sigma^{\mu\nu}
    -R_{\mu\nu}u^\mu u^\nu
    +\overset{(g)}{\nabla}_\mu a^\mu,
\end{align}
where $\theta:=\overset{(g)}{\nabla}_\mu u^\mu$ is the expansion and $\sigma_{\mu\nu}$ is the shear. Related normalized-gradient systems can develop caustics~\cite{Babichev2016,BabichevRamazanov2017a,BabichevRamazanov2017b}. In \MAGIC{}, the spatial gradient of $m$ supplies a geometric acceleration term that can alter the focusing behavior, but the present analysis does not concretely resolve the presence or absence of caustics---this remains an open question.

\section{Fixed volume form and residual symmetry}\label{sec:Ward}

\subsection{Residual diffeomorphism symmetry}\label{Diff}

The action built from Eq.~\eqref{newTerm} is covariant under \emph{passive} coordinate transformations provided that $\mathfrak{m}$ transforms as a weight-$2$ density. This is not the same, however, as an \emph{active} diffeomorphism gauge symmetry.

A \emph{spurionic} transformation, under which $\mathfrak{m}$ is varied together with the dynamical fields, is useful for deriving the fixed-volume-form consistency identity but is not a gauge transformation of a theory with a fixed volume form. Instead, the residual active gauge generators are the vector fields $\zeta^\mu$ satisfying
\begin{align}\label{residual-diff}
  \pounds_\zeta \mathfrak{m}=0,
\end{align}
where $\pounds_\zeta$ denotes the Lie derivative along $\zeta^\mu$. Because $\mathfrak{m}$ is a weight-$2$ density, Eq.~\eqref{residual-diff} reads
\begin{align}\label{Lie-mfrak}
  \zeta^\rho\partial_\rho \mathfrak{m} +2\mathfrak{m}\,\partial_\rho\zeta^\rho=0.
\end{align}
Assuming that $\mathfrak{m}$ is positive, smooth, and nowhere vanishing, Eq.~\eqref{residual-diff} is equivalent to
\begin{align}\label{4.5}
  \partial_\rho \left( \sqrt{\mathfrak{m}}\, \zeta^\rho\right) =0.
 \end{align}
Recalling the fixed volume form in Eq.~\eqref{fixed volume form}, Eqs.~\eqref{residual-diff} and~\eqref{4.5} are equivalent to
\begin{align}\label{weighted-TDiff}
  \pounds_\zeta\boldsymbol{\omega}_{\mathfrak{m}}=0.
\end{align}
The unbroken symmetry is therefore the group of \emph{transverse diffeomorphisms} (TDiffs)
\begin{align}
  \mathrm{TDiff}_{\mathfrak{m}}
    :=
\left\{\varphi\in\mathrm{Diff}(\mathcal M)
    \;\middle|\;
    \varphi^*\boldsymbol{\omega}_{\mathfrak{m}}
    =\boldsymbol{\omega}_{\mathfrak{m}}\right\},
\end{align}
the volume-preserving diffeomorphisms defined by the background volume form selected by $\mathfrak{m}$.

In an $\boldsymbol{\omega}_{\mathfrak{m}}$-adapted coordinate system, say $\boldsymbol{\omega}_{\mathfrak{m}} = \mathrm{d}^4 y$, we recover from Eq.~\eqref{4.5} the familiar TDiff condition $\partial_\rho\zeta^\rho=0$. The statement that a component function $\mathfrak{m}(x)$ is or is not constant is therefore coordinate dependent; in either case, the preserved geometric structure and the three-function local transverse gauge symmetry are unchanged.

Volume-preserving diffeomorphisms are familiar from modified-gravity frameworks~\cite{TDiff2,WeylTDiff,WTDiff}, including unimodular gravity~\cite{UniModular0,UniModular1,UniModular2_Pedagogy,UniModular3}. In \MAGIC{}, the preserved object is specifically $\boldsymbol{\omega}_{\mathfrak{m}}$, and the natural determinant variable is the scalar $\sqrt{-g}/\sqrt{\mathfrak{m}}$.

\subsection{Energy-momentum conservation under $\mathrm{TDiff}_{\mathfrak{m}}$}
\label{sec:TDiff-conservation}

We now derive the energy-momentum identity associated with the residual symmetry identified in Sec.~\ref{Diff}. Under an arbitrary variation of the metric $g_{\mu\nu}$, matter fields $\Psi$, \Moustafa{and prescribed density $\mathfrak m$}, the matter action varies as
\begin{align}\label{matter-general-variation}
\delta S_{\mathrm{m}}
& = \frac{1}{2}
\int \mathrm{d}^4x \sqrt{-g}\,
T^{\mu\nu} \, \delta g_{\mu\nu} \notag\\
&\qquad +\int \mathrm{d}^4x \left(\frac{\delta S_{\mathrm{m}}}{\delta\Psi}
\delta\Psi +\frac{\delta S_{\mathrm{m}}}{\delta \mathfrak{m}}
\delta \mathfrak{m} \right).
\end{align}
Assuming the matter equations of motion hold, that is, $\delta S_{\mathrm{m}} /\delta\Psi =0$, the on-shell variation becomes
\begin{align}\label{matter-onshell-variation}
\delta S_{\mathrm{m}}
=
\frac{1}{2}
\int \mathrm{d}^4x \sqrt{-g}\,
T^{\mu\nu} \,\delta g_{\mu\nu} +\int \mathrm{d}^4x \frac{\delta S_{\mathrm{m}}}{\delta \mathfrak{m}}
\delta \mathfrak{m}.
\end{align}
Under a diffeomorphism generated by $\zeta^\mu$, the metric transforms as
\begin{align}\label{metric-Lie-Levi-Civita}
\delta_\zeta g_{\mu\nu}
=
\pounds_\zeta g_{\mu\nu}
= 2\overset{(g)}{\nabla} _{(\mu}\zeta_{\nu)}.
\end{align}
Substituting Eq.~\eqref{metric-Lie-Levi-Civita} into Eq.~\eqref{matter-onshell-variation}, and using the symmetry of $T^{\mu\nu}$, gives
\begin{align}
\delta_\zeta S_{\mathrm{m}}
&=
\int \mathrm{d}^4x \sqrt{-g}\,
T^{\mu\nu}
\overset{(g)}{\nabla}_{\mu}\zeta_{\nu} +\int \mathrm{d}^4x\frac{\delta S_{\mathrm{m}}}{\delta \mathfrak{m}}
\delta_\zeta \mathfrak{m}
\notag\\
&=
-\int \mathrm{d}^4x \sqrt{-g} \zeta_\nu
\overset{(g)}{\nabla}_{\mu}
T^{\mu\nu} +\int \mathrm{d}^4x \frac{\delta S_{\mathrm{m}}}{\delta \mathfrak{m}}
\delta_\zeta \mathfrak{m},
\label{matter-diff-variation}
\end{align}
where a boundary term has been discarded in the second equality after integrating by parts. A fully diffeomorphism-invariant matter action has no $\mathfrak{m}$-dependence, $S_\mathrm{m} = S_\mathrm{m}[\mathbf{g},\Psi]$, and hence $\delta S_{\mathrm{m}} /\delta \mathfrak{m} =0$. Moreover, in that case the generator $\zeta^\mu$ is arbitrary, so the vanishing of Eq.~\eqref{matter-diff-variation} implies
\begin{align}\label{matterConservation}
\overset{(g)}{\nabla}_{\mu}T^{\mu\nu}
=0
\qquad
(\mathrm{full~Diff}).
\end{align}

The result is weaker, however, if the matter action is \emph{not} fully diffeomorphism-invariant, $S_\mathrm{m} = S_\mathrm{m}[\mathbf{g},\Psi; \mathfrak{m}]$. Fundamentally, our theory is a toy model of a quantum-gravitational effect, and, just as matter couples to classical gravity, it must also couple to quantum gravity. A dependence on $\mathfrak{m}$ may therefore arise in the matter sector after ultraviolet degrees of freedom associated with chronology protection are integrated out, leaving ${\mathfrak{m}}$ as a fixed infrared remnant.

For a residual \emph{active} diffeomorphism, $\delta_\zeta\mathfrak{m}=\pounds_\zeta\mathfrak{m}=0$ by Eq.~\eqref{residual-diff}. The generator $\zeta^\mu$ is therefore not arbitrary; it is restricted by Eq.~\eqref{4.5}, which locally has the general solution
\begin{align}\label{TDiff-generator-potential}
\sqrt{\mathfrak{m}}\,\zeta^\mu
=
\partial_\nu \mathfrak{B}^{[\nu\mu]},
\end{align}
where $\mathfrak{B}^{[\nu\mu]}$ is an arbitrary antisymmetric tensor density. Substituting Eq.~\eqref{TDiff-generator-potential} into Eq.~\eqref{matter-diff-variation} and using Eq.~\eqref{ratio} to trade the density $\mathfrak{m}$ for the scalar $m$, we find
 \begin{align}
  0 = \delta_\zeta S_{ \mathrm{m}}  &= - \int\mathrm{d}^4x \,
\frac{1}{\sqrt{m}}\,
\Big(\overset{(g)}{\nabla}_{\rho}T^\rho{}_\nu\Big) \,
\partial_\mu \mathfrak{B}^{[\mu\nu]}
\notag\\
&= \int\mathrm{d}^4x \,
\partial_\mu \left(\frac{1}{\sqrt{m}}\,
\overset{(g)}{\nabla}_{\rho}T^\rho{}_\nu \right) \mathfrak{B}^{[\mu\nu]},
\label{TDiff-Ward-before-antisym}
\end{align}
where a boundary term has again been discarded after integrating by parts. Since $\mathfrak{B}^{[\mu\nu]}$ is arbitrary and antisymmetric, only the antisymmetric part of the quantity multiplying it contributes. The local Noether identity is therefore
 \begin{align}  \label{TDiff-Ward-explicit}
\partial_{[\mu}
 \left(\frac{1}{\sqrt{m}}\,
\overset{(g)}{\nabla}_{\rho}T^\rho{}_{\nu]}\right) =0.
 \end{align}
Thus, $\overset{(g)}{\nabla}_{\mu}T^\mu{}_\nu$ is not required to vanish under $\mathrm{TDiff}_{\mathfrak{m}}$. Instead, the one-form
\begin{align}
\boldsymbol{K}:= \frac{1}{\sqrt{m}}\,
\overset{(g)}{\nabla}_{\mu}T^\mu{}_{\nu}\, \mathrm{d}x^\nu
\end{align}
must be closed. Locally, the Poincar\'e lemma implies the existence of a scalar field $\psi$ such that $\boldsymbol{K} = \mathrm{d} \psi$, or
\begin{align}\label{TDiff-local-potential}
\frac{1}{\sqrt{m}}\,
\overset{(g)}{\nabla}_{\rho}T^\rho{}_{\nu}
=
\partial_\nu \psi.
\end{align}
Hence, the most general local energy-momentum identity under $\mathrm{TDiff}_{\mathfrak{m}}$ is
\begin{align}\label{TDiff-conservation-law}
\overset{(g)}{\nabla}_{\mu}T^\mu{}_\nu
= \sqrt{m}\,
\partial_\nu \psi \qquad
(\mathrm{TDiff}_{\mathfrak{m}}).
\end{align}
Within the non-degenerate branch $m>0$, so exact conservation is recovered only when $\psi$ is constant.

\Moustafa{Given a matter action $S_{\mathrm{m}}$, the scalar field $\psi$ may be constructed explicitly---one then obtains a \emph{consistency equation} compatible with Eq.~\eqref{TDiff-conservation-law}. If the matter action is not specified, however, such a consistency equation becomes a constraint on the class of allowed actions. This identification follows by formally retaining the last term in Eq.~\eqref{matter-diff-variation}.}

\Moustafa{Consider the simultaneous transformation of all arguments of $S_{\mathrm m}[\mathbf g,\Psi;\mathfrak m]$. In this calculation the generator $\zeta^\nu$ is arbitrary, and $\mathfrak{m}$ transforms as
\begin{align}\label{matter-background-Lie}
\delta_\zeta\mathfrak m
=\zeta^\nu\partial_\nu\mathfrak m
+2\mathfrak m\,\partial_\nu\zeta^\nu.
\end{align}
Setting $\delta_\zeta S_{\mathrm{m}} =0$ in Eq.~\eqref{matter-diff-variation} and rearranging, one obtains
\begin{align}\label{matter-spurion-variation}
\int\mathrm d^4x\,\sqrt{-g} \zeta_\nu
\overset{(g)}{\nabla}_{\mu}T^{\mu\nu} =\int \mathrm{d}^4x \left(\zeta^\nu\partial_\nu\mathfrak m
+2\mathfrak m\,\partial_\nu\zeta^\nu\right) \frac{\delta S_{\mathrm m}}{\delta\mathfrak m}
.
\end{align}
Integrating by parts and using the arbitrariness of $\zeta^\nu$,
\begin{align}\label{matter-background-Ward}
\sqrt{-g}\,
\overset{(g)}{\nabla}_{\mu}T^\mu{}_{\nu}
={}&\frac{\delta S_{\mathrm m}}{\delta\mathfrak m}
\partial_\nu\mathfrak m
-2\partial_\nu\left(
\mathfrak m\frac{\delta S_{\mathrm m}}{\delta\mathfrak m}
\right).
\end{align}
Using $m=\mathfrak m/(-g)$, we can write this as
\begin{align}\label{matter-background-gradient}
\overset{(g)}{\nabla}_{\mu}T^\mu{}_{\nu}
=\sqrt m\,\partial_\nu\left(-2
\sqrt{\mathfrak m}\frac{\delta S_{\mathrm m}}{\delta\mathfrak m}
\right).
\end{align}
This is precisely the form in Eq.~\eqref{TDiff-conservation-law}, with (up to an additive constant)
\begin{align}\label{matter-psi-identification}
\psi=-2\sqrt{\mathfrak m}\,
\frac{\delta S_{\mathrm m}}{\delta\mathfrak m}.
\end{align}

Thus, the $\mathrm{TDiff}_{\mathfrak{m}}$ argument establishes the local existence of $\psi$, while the simultaneous-transformation identity constructs $\psi$ in terms of the matter action. The latter uses formal covariance under transformations of all arguments. Again, if the matter action is independent of $\mathfrak m$, Eq.~\eqref{matter-background-gradient} reduces to ordinary covariant conservation.}

\subsection{Energy-momentum conservation in MAGIC}
\label{sec:MAGIC-conservation}

We now specialize the general identity in Eq.~\eqref{TDiff-conservation-law} to the \MAGIC{} field equations. Applying $\overset{(g)}{\nabla}{}^\mu$ to Eq.~\eqref{NewEinstein} gives
\begin{align}\label{Bianchi-source-balance}
\overset{(g)}{\nabla}{}^\mu T_{\mu\nu}
=
\overset{(g)}{\nabla}{}^\mu\mathcal{C}_{\mu\nu}.
\end{align}
To calculate the right-hand side, apply $\overset{(g)}{\nabla}{}^\mu$ to the definition of $\mathcal{C}_{\mu\nu}$ in Eq.~\eqref{Delta-QG-reduced}:
\begin{align}
  \overset{(g)}{\nabla}{}^\mu
\mathcal{C}_{\mu\nu}
&=\overset{(g)}{\nabla}{}^\mu
 \left(\lambda\partial_\mu\phi\partial_\nu\phi
 \right) +
2\partial_\nu
\left(\lambda m^2
 \right)
 \notag\\
&=
 \Big(\overset{(g)}{\nabla}{}^\mu
\left(\lambda\partial_\mu\phi\right)\Big)
\partial_\nu\phi +\lambda \overset{(g)}{\nabla}{}^\mu \phi
\overset{(g)}{\nabla}_{\mu}\partial_\nu\phi \notag\\
&~~~+
2\partial_\nu
\left(\lambda m^2
\right).
\label{M-divergence-expanded}
\end{align}
The first term in the second line of Eq.~\eqref{M-divergence-expanded} vanishes on account of Eq.~\eqref{phiConservation}. Moreover, since $\phi$ is a scalar,
 \begin{align}
  \overset{(g)}{\nabla}_{\mu}\partial_\nu\phi
=
\overset{(g)}{\nabla}_{\nu}\partial_\mu\phi.
\end{align}
Consequently,
\begin{align}
\overset{(g)}{\nabla}{}^\mu \phi\,
\overset{(g)}{\nabla}_{\mu}\partial_\nu\phi =
\overset{(g)}{\nabla}{}^\mu \phi\,
\overset{(g)}{\nabla}_{\nu}\partial_\mu\phi =
\frac{1}{2}
\partial_\nu
\left(
 g ^{\alpha\beta}
\partial_\alpha \phi \partial_\beta \phi
 \right).
\label{scalar-gradient-identity}
\end{align}
Using the norm constraint in Eq.~\eqref{7.13}, we obtain
\begin{align}
\overset{(g)}{\nabla}{}^\mu \phi\,
\overset{(g)}{\nabla}_{\mu}\partial_\nu\phi
 =
  -\frac{1}{2}\partial_\nu m^2.
 \end{align}
Equation~\eqref{M-divergence-expanded} thus reduces to
 \begin{align}  \label{M-divergence}
\overset{(g)}{\nabla}{}^\mu
\mathcal{C}_{\mu\nu} =
  -\frac{\lambda}{2}\partial_\nu m^2
+
2\partial_\nu
\left(
\lambda m^2
\right) =\sqrt{m}\, \partial_\nu \left(2
\lambda m^{3/2}
\right).
 \end{align}
Substituting back into Eq.~\eqref{Bianchi-source-balance} gives
\begin{align}\label{MAGIC-conservation-equation}
\overset{(g)}{\nabla}{}^\mu T_{\mu\nu} =\sqrt{m}\, \partial_\nu \left(2
\lambda m^{3/2}
\right).
\end{align}
This indeed takes the form permitted by $\mathrm{TDiff}_{\mathfrak{m}}$ in Eq.~\eqref{TDiff-conservation-law} with (again up to an additive constant)
\begin{align}
\psi_{\mathrm{\MAGIC{}}}
=
2\lambda m^{3/2}.
\end{align}
Hence, exact energy-momentum conservation in the matter sector of \MAGIC{} requires $\lambda \propto m^{-3/2}$.

\Moustafa{Comparing Eq.~\eqref{matter-background-gradient} with the \MAGIC{} expression Eq.~\eqref{MAGIC-conservation-equation} therefore yields
\begin{align}\label{consistencyExchange}
    \partial_\mu\left(
\lambda m^{3/2}
+\sqrt{\mathfrak m}\,\frac{\delta S_{\mathrm m}}{\delta\mathfrak m}
\right)=0.
\end{align}
This is the consistency equation mentioned above. The phenomenological cosmology below is conditional on a matter action realizing both the assumed equation of state and this identity.
}

\section{Spatially homogeneous and isotropic cosmology}\label{sec:cosmology}

In this section, we consider homogeneous and isotropic solutions governed by the \MAGIC{}-modified gravitational equations. In coordinates $(\tau,r,\theta,\varphi)$, we take the metric to have the Friedmann--Lema{\^i}tre--Robertson--Walker (FLRW) form
 \begin{align}
    \mathrm{d}s^2 = - \mathrm{d}\tau^2 + a^2(\tau)\left[\frac{\mathrm{d}r^2}{1-k r^2} + r^2 \mathrm{d}\Omega^2\right],
\end{align}
where $\mathrm{d}\Omega^2 := \mathrm{d}\theta^2 +\sin^2{\theta} \mathrm{d}\varphi^2$ is the metric on the round 2-sphere, $a(\tau)$ is the scale factor, $\tau$ is cosmic time, and $k\in\mathbb{R}$ is the constant spatial-curvature parameter. The Einstein tensor is diagonal, and its nonvanishing components are
\begin{align}
    G_{\tau\tau}
    &=3\left( H^2 +\frac{k}{a^2}\right),
    \\
    G_{rr}
    &=-\frac{a^2}{1-k r^2}
    \left[2  \dot{H}  + 3H^2 +\frac{k}{a^2}\right],
    \\
    G_{\theta\theta}
    &=-a^2 r^2\left[2  \dot{H}  + 3H^2 +\frac{k}{a^2}
    \right],
\end{align}
with $G_{\varphi\varphi} = \sin^2{\theta} \,G_{\theta\theta}$ by spherical symmetry. Here dots denote derivatives with respect to \(\tau\), and the \emph{Hubble parameter} is
\begin{align}
  H(\tau) := \frac{\dot{a}(\tau)}{a(\tau)}.
\end{align}

Spatial homogeneity and isotropy require the scalar field $\phi$ to depend only on cosmic time. The normalization condition, Eq.~\eqref{7.13}, then reduces to
\begin{align}\label{eq:phi_norm_FLRW}
    \dot{\phi}(\tau) = m(\tau),
\end{align}
up to an overall sign.

\subsection{Modified Friedmann equations}

We take the ordinary-matter energy-momentum tensor to have the perfect-fluid form,
\begin{align}
  T_{\mu\nu} = \left(\varepsilon +p\right) u_\mu u_\nu + p\, g_{\mu \nu},
\end{align}
with $u_\mu$ given by Eq.~\eqref{unitTimelike}; \Moustafa{in this case it is simply 
\begin{align}\label{u_mu}
    u_\mu = -\delta^\tau_\mu.
\end{align}} The $\tau\tau$ component of Eq.~\eqref{NewEinstein} then gives
\begin{align} \label{friedmann-rho}
     3\left( H^2 +\frac{k}{a^2}\right) -\Lambda  - \lambda m^2 = \varepsilon,
\end{align}
while the spatial components give
 \begin{align} \label{acceleration-rho}
     2 \dot H + 3H^2 +\frac{k}{a^2}-\Lambda -2 \lambda m^2 = -p.
\end{align}
Equivalently, eliminating $H^2$ from Eq.~\eqref{acceleration-rho},
 \begin{align}
  \dot H
 = \frac{k}{a^2}+\frac{\lambda m^2}{2} - \frac{1}{2} \left(  \varepsilon  + p\right).
    \label{eq:mod_Friedmann_1_tau}
 \end{align}
  
For consistency, Eq.~\eqref{connection-integrability-Q} must also be satisfied. For FLRW, Eq.~\eqref{source} gives the projective current
 \begin{align}\label{7.11}
  \mathfrak{J}^\mu =  - \sqrt{-g}\,\lambda m\, \delta^\mu_\tau.
 \end{align}
Now, the metric determinant is $\sqrt{-g}=a^3\sqrt{\gamma}$, where $\sqrt{\gamma}:= r^2 \sin{\theta} / \sqrt{1 -kr^2}$ is independent of $\tau$, and hence Eq.~\eqref{connection-integrability-Q} yields
 \begin{align}  \label{lambda-m-FLRW}
  \lambda m\, a^3 = C,
 \end{align}
where $C$ is an integration constant. By construction, both $m$ and $a$ are positive; the sign of $\lambda$ is therefore determined by the sign of $C$. \Moustafa{Combining Eqs.~\eqref{7.11} and~\eqref{lambda-m-FLRW}, the projective current reduces to
\begin{align}\label{projective_FLRW}
    \mathfrak{J}^\mu =  - C \sqrt{\gamma}\, \delta^\mu_\tau
\end{align}}

\Moustafa{Another consequence of Eq.~\eqref{connection-integrability-Q} is that the nonzero homogeneous branch $C\neq0$ cannot be realized with a globally smooth tensor multiplier on a compact boundaryless slice $\Sigma$. To see this, contract Eq.~\eqref{Maxwell_eq} with the unit normal to $\Sigma$, that is $u_\mu$, and integrate the result on such a slice. One has
\begin{align}
    \int_{\Sigma} \mathrm{d}^3 x\,  u_\mu \partial_\nu \mathfrak{F}^{\mu\nu} = \int_{\Sigma} \mathrm{d}^3 x\,  u_\mu \mathfrak{J}^\mu.
\end{align}
Using Eqs.~\eqref{u_mu} and~\eqref{projective_FLRW} together with the antisymmetry of $\mathfrak{F}^{\mu\nu}$, this becomes
\begin{align}
    \int_{\Sigma} \mathrm{d}^3 x\,  \partial_i \mathfrak{F}^{\tau i} = -C \int_{\Sigma} \mathrm{d}^3 x \sqrt{\gamma}.
\end{align}
The right-hand side of the equation is simply $-C\operatorname{Vol}(\Sigma)$. The left-hand side can be re-written using Gauss's theorem as an integral over the boundary $\partial \Sigma$. Thus,
\begin{align}
    \int_{\partial\Sigma} \mathrm{d}^2 x\,  n_i \mathfrak{F}^{\tau i} = -C \operatorname{Vol}(\Sigma),
\end{align}
where $n_i$ is the normal to $\partial\Sigma$. For a compact boundaryless slice, $\partial\Sigma=\varnothing$, so the left-hand side vanishes. Since the volume of $\Sigma$ is nonzero, one obtains
\begin{align}
    C=0, \qquad \text{for compact boundaryless } \Sigma.
\end{align}
We therefore restrict the nonzero homogeneous branch studied below to noncompact slices.

For example, a nonzero branch can be realized on a flat slice $\Sigma=\mathbb R^3$ with $k=0$. In Cartesian comoving coordinates, a corresponding smooth solution is
\begin{align}
 \mathfrak F^{\tau i}=-\frac C3x^i,\qquad
 \mathfrak F^{ij}=0.
\end{align}
This satisfies every component of Eq.~\eqref{Maxwell_eq}, with flux $-4\pi C R^3/3$ through a comoving sphere of radius $R$. We allow this non-decaying $\mathfrak{F}^{\mu\nu}$ and use compactly supported variations for the bulk variational problem. Note that this does not contradict our assumed homogeneity; homogeneity refers to the geometry and effective source, not the auxiliary representative chosen above.}

\Moustafa{With that caveat in mind, we now substitute} Eq.~\eqref{lambda-m-FLRW} back into Eqs.~\eqref{friedmann-rho} and~\eqref{eq:mod_Friedmann_1_tau} to eliminate $\lambda$. We obtain the final version of the modified Friedmann equations:
 \begin{align}
  H^2
 &= - \frac{k}{a^2}  + \frac{1}{3} \left( \varepsilon  + \Lambda +\frac{C m}{a^3}\right),
    \label{Friedmann_1} \\
  \dot H
 &= \frac{k}{a^2} -\frac{1}{2}\left( \varepsilon  +p -\frac{C m}{a^3}\right).
\label{Friedmann_2}
 \end{align}

\subsection{Energy exchange with matter}

The chronology correction thus appears as an effective source with
 \begin{align}  \label{effective-fluid-FLRW}
     \varepsilon_{\mathrm{eff}}  = \frac{C m}{a^3}, \qquad p_{\mathrm{eff}} = -\frac{2C m}{a^3},
\end{align}
and Eq.~\eqref{MAGIC-conservation-equation} gives
 \begin{align}  \label{nonCons}
    \dot{\varepsilon}  + 3H\left(\varepsilon + p  \right) = \mathcal{Q},
\end{align}
where
\begin{align}\label{Q-FLRW}
  \mathcal{Q} := -2 \sqrt{m}\, \frac{\mathrm{d}}{\mathrm{d}\tau} \left(\frac{\varepsilon_{\mathrm{eff}}}{\sqrt{m}}\right).
\end{align}

Equation~\eqref{nonCons} is the homogeneous Bianchi consistency condition derived in Sec.~\ref{sec:Ward}. It permits effective energy exchange, \Moustafa{provided that a matter action realizes the consistency condition in Eq.~\eqref{consistencyExchange}.} A positive $\mathcal Q$ corresponds to energy transfer into matter \Moustafa{through the assumed effective coupling}. As we shall show in the next subsection, an assumed profile for $m$ can make this effect large in the early Universe and negligible at late times.

Without ultraviolet input, the model does not determine either the profile of $m$ or its coupling to matter. We therefore restrict the following discussion to an illustrative profile

\subsection{A toy early-Universe example}
\label{sec:early-universe-example}

As a pedagogical illustration, consider a homogeneous fixed volume form and a solution branch on which the scalar ratio $m$ obeys, during a finite early epoch, the power law
\begin{align}
    m(a)
    =m_0\left(\frac{a_0}{a}\right)^n,
    \qquad n>0.
    \label{illustrative-m-profile}
\end{align}
The restriction $n>0$ ensures that $m$ diverges upon approaching the would-be degenerate limit $a\to0$. Defining
\begin{align}
    {\varepsilon_{\mathrm{eff}}}_0
    :=\frac{Cm_0}{a_0^3},
    \qquad C>0,
\end{align}
Eqs.~\eqref{effective-fluid-FLRW} and~\eqref{Q-FLRW} give
\begin{align}
    \varepsilon_{\mathrm{eff}}
    &=
    {\varepsilon_{\mathrm{eff}}}_0
    \left(\frac{a_0}{a}\right)^{n+3},
    \label{illustrative-effective-density}
    \\
    \mathcal Q
    &=
    (n+6)H{\varepsilon_{\mathrm{eff}}}_0
    \left(\frac{a_0}{a}\right)^{n+3}.
    \label{illustrative-pump}
\end{align}
Thus, on an expanding branch, the effective sector transfers energy into matter.

Suppose that the effective $\mathrm{TDiff}_{\mathfrak{m}}$-invariant matter sector is described at the homogeneous level by a general barotropic perfect fluid,
\begin{align}
    p=w\varepsilon,
\end{align}
where $w$ is constant. Equation~\eqref{nonCons} then becomes
\begin{align}
    \dot{\varepsilon}
    +3H(1+w)\varepsilon
    =
    (n+6)H{\varepsilon_{\mathrm{eff}}}_0
    \left(\frac{a_0}{a}\right)^{n+3}.
    \label{illustrative-fluid-continuity}
\end{align}
On any interval over which $a$ is monotonic, this may be written as
\begin{align}
    a\frac{\mathrm{d}\varepsilon}{\mathrm{d}a}
    +3(1+w)\varepsilon
    =
    (n+6){\varepsilon_{\mathrm{eff}}}_0
    \left(\frac{a_0}{a}\right)^{n+3}.
\end{align}
For $n\neq3w$, the solution is
\begin{align}
    \varepsilon
    &=
    \left(\frac{a_0}{a}\right)^{3(1+w)}
    \Bigg[
        \varepsilon_0
        +\frac{n+6}{n-3w}{\varepsilon_{\mathrm{eff}}}_0
        \left(
            1-
            \left(\frac{a_0}{a}\right)^{n-3w}
        \right)
    \Bigg],
    \label{illustrative-fluid-solution}
\end{align}
where $\varepsilon_0 := \varepsilon(a_0)$. The value $n=3w$ is resonant because the chronology-sector density then has the same scale-factor dependence as the homogeneous perfect-fluid solution. In this case, the solution is instead
\begin{align}
    \varepsilon
    &=
    \left(\frac{a_0}{a}\right)^{3(1+w)}
    \left[
        \varepsilon_0
        +(n+6){\varepsilon_{\mathrm{eff}}}_0
        \ln\left(\frac{a}{a_0}\right)
    \right].
    \label{illustrative-resonant-solution}
\end{align}
Representative cases are summarized in Table~\ref{tab:illustrative-fluid-examples}.
\begin{table*}[t]
\caption{
Representative solutions of Eq.~\eqref{illustrative-fluid-continuity}.
}
\label{tab:illustrative-fluid-examples}
\renewcommand{\arraystretch}{1.5}
\begin{ruledtabular}
\begin{tabular}{lccc}
Matter or profile
&
$w$
&
$n$
&
$\varepsilon(a)$
\\
\hline
Dust
&
$0$
&
$2$
&
$\displaystyle
(\varepsilon_0+4{\varepsilon_{\mathrm{eff}}}_0)(a_0/a)^3-4{\varepsilon_{\mathrm{eff}}}_0 (a_0/a)^5
$
\\
Radiation
&
$1/3$
&
$2$
&
$\displaystyle
(\varepsilon_0+8{\varepsilon_{\mathrm{eff}}}_0) (a_0/a)^4-8{\varepsilon_{\mathrm{eff}}}_0 (a_0/a)^5
$
\\
Resonant radiation
&
$1/3$
&
$1$
&
$\displaystyle
(a_0/a)^4
\Big[
\varepsilon_0 +7{\varepsilon_{\mathrm{eff}}}_0 \ln{(a/a_0)}
\Big]
$
\\
Generic
&
$w \neq 2$
&
$6$
&
$\displaystyle
\left[
\varepsilon_0 +\frac{4{\varepsilon_{\mathrm{eff}}}_0}{2-w}
\right] (a_0/a)^{3(1+w)}
-\frac{4{\varepsilon_{\mathrm{eff}}}_0}{2-w} (a_0/a)^9
$
\end{tabular}
\end{ruledtabular}
\end{table*}

For $n>3w$, the chronology correction decays faster than the ordinary perfect-fluid contribution while leaving a finite enhancement of its comoving energy. Requiring $\varepsilon\geq0$ restricts the solution to
\begin{align}
    \frac{a}{a_0}
    \geq
    \left[
        \frac{
            (n+6){\varepsilon_{\mathrm{eff}}}_0
        }{
            (n-3w)\varepsilon_0
            +(n+6){\varepsilon_{\mathrm{eff}}}_0
        }
    \right]^{1/(n-3w)}.
    \label{illustrative-positive-domain}
\end{align}
The region below this bound is not part of the positive-energy branch. At $n=3w$, the corresponding enhancement is logarithmic, whereas for $0<n<3w$ the chronology correction decays more slowly than the homogeneous perfect-fluid contribution and is therefore not naturally confined to the early Universe unless the assumed profile for $m(a)$ terminates after a finite epoch.

At the homogeneous level, this resembles open-system cosmological matter and entropy production~\cite{Prigogine1988MatterCreation,Calvao1992MatterCreation}. A transient source of this kind could in principle modify the pre-nucleosynthesis thermal history and relic production~\cite{Giudice2001LargestTemperature}. However, since neither Eq.~\eqref{illustrative-m-profile} nor the microscopic thermalization channel is fixed by the present \MAGIC{} action, and since the scalar branch---as currently formulated---is gradient-unstable as discussed in Sec.~\ref{sec:perturbations}, this example should be regarded only as a \Moustafa{conditional phenomenological background solution; a matter action realizing both the assumed equation of state and the consistency Eq.~\eqref{consistencyExchange} has not been specified}.

\section{Local scalar representation of \MAGIC{} and perturbative spectrum}\label{sec:perturbations}

\subsection{Equivalent local scalar representation}
On a contractible patch, the closure equation $R'_{\mu\nu}=0$ makes the non-metricity one-form exact. After the Palatini connection equation is solved, the projective shift leaves the Einstein--Hilbert action unchanged. The local metric and scalar equations can therefore be obtained equivalently from
 \begin{align}  \label{local-scalar-action}
    &S_{\mathrm{loc}}[\mathbf{g},\Psi,\phi,\lambda;\mathfrak{m}]
 = S_{\mathrm{EH}}[\mathbf{g}] +S_{\mathrm{m}}[\mathbf{g},\Psi; \mathfrak{m}]\notag \\
    &~~~+\frac{1}{2}\int \mathrm{d}^4x\sqrt{-g}\,\lambda
\Big(g^{\mu\nu}\partial_\mu\phi\partial_\nu\phi
    +m^2(g;\mathfrak{m})\Big) .
\end{align}
Here $S_{\mathrm{EH}}[\mathbf{g}]$ is the \emph{metric} Einstein--Hilbert action,
 \begin{align}
  S_{\mathrm{EH}}[\mathbf{g}] := \frac{1}{2}\int \mathrm{d}^4x \sqrt{-g}
\left(
  R[\mathbf{g}] - 2\Lambda
\right).
 \end{align}
 
This is a \emph{local} reduction of \MAGIC{}, not a replacement for its affine definition. Because $m$ depends on the metric determinant, metric variation of Eq.~\eqref{local-scalar-action} reproduces the correction tensor $\mathcal{C}_{\mu\nu}$, Eq.~\eqref{Delta-QG-reduced}. Variation with respect to $\phi$ gives the projective current integrability condition, Eq.~\eqref{phiConservation}. \Moustafa{A \emph{global} lift additionally requires a smooth solution of Eq.~\eqref{Maxwell_eq} with the chosen topology and boundary data, as discussed in Sec.~\ref{sec:cosmology}.}

\subsection{Relation to Generalized Unimodular Gravity}

We may write the metric in ADM form,
\begin{align}
  \mathrm{d}s^2
 = -N^2\mathrm{d}t^2
 + h_{ ij }  \left(\mathrm{d}x^i+N^i\mathrm{d}t\right)
              \left( \mathrm{d}x^j + N^j\mathrm{d}t \right),
 \end{align}
with $h:=\det h_{ij}$ and $-g=N^2h$. Since the time gradient is everywhere timelike, we may use $\phi$ as a local time coordinate. In unitary gauge, $\phi=t$, and hence the norm constraint becomes
 \begin{align}
  -\frac{1}{N^2}
 + \frac{\mathfrak{m}^2}{N^4h^2} =0 ,
\end{align}
so, on the future-directed branch,
 \begin{align}  \label{GUMG-constraint}
  N = \frac{\mathfrak{m}}{h}.
 \end{align}
In coordinates adapted to the fixed volume form, the nonvanishing density component $\mathfrak{m}$ is a constant and can be set to unity without loss of generality. Equation~\eqref{GUMG-constraint} then has the form $N=f(h)$ with
 \begin{align}  \label{8.6}
  \ln f(h) = - \ln h.
\end{align}
Generalized unimodular gravity~\cite{GUMGDynamics,GUMGInflation,GUMGKessence} with $N=f(h)$ admits a formal perfect-fluid parametrization with
 \begin{align}  \label{GUMG-w}
  w_{\mathrm{eff}} = 2\frac{\mathrm{d}\ln f}{\mathrm{d}\ln h},
\end{align}
so Eq.~\eqref{8.6} gives $w_{\mathrm{eff}}=-2$, independently reproducing the formal parameter in Eq.~\eqref{density_pressure}. Thus the local scalar dynamics of \MAGIC{} contains the $w_{\mathrm{eff}}=-2$ generalized-unimodular branch; the metric--affine construction supplies its projective origin and chronology interpretation.

\subsection{Scalar and tensor quadratic actions}

To determine the spectrum fixed by the chronology sector without specifying a microscopic matter completion, we set $S_{\mathrm m}=0$ \Moustafa{and $\Lambda=0$} in this subsection and study the matter-free local gravitational sector. Generic TDiff matter contributes its own perturbations and mixing, which can be analyzed only after an explicit $S_{\mathrm m}[\mathbf{g},\Psi;\mathfrak m]$ is chosen.

\Moustafa{Also, we use the non-compact flat realization above on the nontrivial branch $C>0$, so $\varepsilon_{\mathrm{eff}}=\lambda m^2=3H^2>0$. In this matter-free sector, Eq.~\eqref{MAGIC-conservation-equation} gives $\lambda m^{3/2}=\mathrm{const}$; together with Eq.~\eqref{lambda-m-FLRW}, this requires $m\propto a^6$. The background therefore differs from the matter-exchanging example in Sec.~\ref{sec:early-universe-example}.}

For the matter-free, spatially flat, homogeneous branch, the scalar quadratic action is known in canonical generalized-unimodular form~\cite{GUMGDynamics,GUMGInflation,GUMGKessence}. \Moustafa{On constant-$\phi$ slices, define
\begin{align}
    \delta h_{ij}=a^2(-2\chi\delta_{ij}+2\partial_i\partial_j E)
\end{align}
before eliminating the constrained variables. In conformal time $\mathrm d\eta=\mathrm d\tau/a$, with primes denoting $\eta$ derivatives,} it may be written as
 \begin{align}  \label{scalar-quadratic}
  S_{\mathrm{s}}^{(2)}
 =  \frac12\int \mathrm{d}\eta\,\mathrm{d}^3x
\left[
  (v')^2
        +c_s^2 v\,\Delta v
        +\frac{\mathcal Z''}{\mathcal Z}v^2
\right],
 \end{align}
where $v=\mathcal Z \chi$ is the canonically normalized scalar metric perturbation, $\Delta:=\delta^{ij}\partial_i\partial_j$ is the spatial Laplacian, and
 \begin{align}
  \mathcal Z^2
& = 3a^2\frac{\Omega}{w_{\mathrm{eff}}},\\
  c_s^2
 &= \frac{w_{\mathrm{eff}} (1+w_{\mathrm{eff}})}{\Omega}, \\
  \Omega
 &:= 1 + w_{\mathrm{eff}} +2\frac{\mathrm{d}\ln |w_{\mathrm{eff}}|}{\mathrm{d}\ln h}.
\end{align}
For the \MAGIC{} branch, $w_{\mathrm{eff}}=-2$ is constant, and hence
 \begin{align}  \label{MAGIC-perturbation-parameters}
  \Omega = -1,
    \qquad
    \mathcal Z^2=  \frac32a^2,
    \qquad
    c_s^2 = -2.
 \end{align}
The positive value of $\mathcal Z^2$ shows that the scalar mode is not a ghost at quadratic order. Its sound speed squared is negative, however. For a Fourier mode whose wavelength is well below the background curvature scale, Eq.~\eqref{scalar-quadratic} gives
 \begin{align}
  v_k''-2k^2v_k \approx 0,
\end{align}
so the scalar sector has a short-wavelength gradient instability. Analogous instabilities in constrained scalar and other modified-gravity effective theories have been addressed by enlarging the operator content, adding controlled higher-spatial-derivative corrections to the dispersion relation, or embedding the unstable low-energy description in a healthier theory~\cite{HiranoMimeticStability,ZhengMimeticStability,GorjiScordatura,AokiProcaPathologies}. \Moustafa{Stabilization is required before a viable cosmological application can be established; its implementation is left to future work.}

Tensor perturbations are traceless and do not perturb $h$ when parameterized by $h_{ij}=a^2(e^{\boldsymbol{\gamma}})_{ij}$ with $\gamma_i{}^i=0$. The determinant constraint therefore leaves their quadratic action unchanged from GR:
\begin{align}\label{tensor-quadratic}
  S_{\mathrm{T}}^{(2)}
 = \frac{1}{8}\int \mathrm{d}\tau\,\mathrm{d}^3x\,a^3
 \left[
  \dot\gamma_{ij}\dot\gamma^{ij}
        -\frac{1}{a^2}
        \partial_k\gamma_{ij}\partial^k\gamma^{ij}
    \right].
\end{align}
The two tensor polarizations are consequently massless, luminal, and free of ghosts and gradient instabilities at quadratic order on this background.

\subsection{The multiplier and projective sectors}

The perturbative analysis also clarifies the role of $\xi^{[\mu\nu]}$. At quadratic order, its coupling to the projective connection is linear in the homothetic curvature. Variation with respect to $\xi^{[\mu\nu]}$ imposes the closure of the projective one-form, while the reducible transformation in Eqs.~\eqref{xi-gauge}--\eqref{xi-reducibility} removes the corresponding gauge redundancy. Neither $\xi^{[\mu\nu]}$ nor the transverse part of the projective connection acquires a quadratic kinetic operator. Instead, the projective connection equation supplies the current integrability condition that becomes the scalar equation following from Eq.~\eqref{local-scalar-action}.

After the constraints imposed by the lapse, shift, $\lambda$, and $\xi^{[\mu\nu]}$ are enforced, the matter-free local gravitational quadratic FLRW spectrum consists of the two tensor polarizations and the single scalar mode $v$. In this reduction, neither $\xi^{[\mu\nu]}$ nor the transverse projective sector carries a local kinetic mode. Global or branch-dependent sectors may remain, as is familiar in constrained generalized-unimodular systems~\cite{GUMGCanonical,GUMGDynamics}; \Moustafa{the spectrum established here is local and restricted to the stated non-compact positive-density background.}

\section{Summary, Discussion, and Future Directions}\label{sec:discussion}

We have introduced \MAGIC{}, a toy metric--affine model that dynamically implements a timelike, closed non-metricity one-form from the affine connection without introducing additional matter fields. The model is motivated by Hawking's chronology protection conjecture and by the possibility that an ultraviolet mechanism suppressing non-stably causal spacetimes in the gravitational path integral may leave infrared remnants in the effective gravitational action. Thus, \MAGIC{} fits naturally within the broader logic of quantum-gravity phenomenology---one postulates a qualitative ultraviolet effect and studies its infrared consequences.

The \MAGIC{} action implements two complementary constraints. The first is degeneracy-sensitive and, \Moustafa{under regularity and orientation assumptions, yields the classical nondegeneracy result of Eq.~\eqref{eq:5.7}.} The second sets the homothetic curvature to zero, forcing the timelike non-metricity one-form $Q_\mu$ to be closed and hence locally expressible as the gradient of a scalar field, $Q_\mu=2\partial_\mu\phi$. When the resulting closed one-form is globally exact, $\phi$ is a global time function and stable causality follows. The preferred chronology structure is thus extracted from the affine connection rather than introduced as an additional matter field.

Geometrically, this mechanism is understood as the breaking of projective invariance. A one-form direction that is pure gauge in metric--affine GR becomes constrained geometric structure, fixing the torsion and non-metricity of the selected connection. The resulting correction $\mathcal{C}_{\mu\nu}$ is therefore most naturally placed on the geometric side of the Einstein equations.

The necessary existence of a fixed volume form $\boldsymbol{\omega}_{\mathfrak{m}}$ for this approach to chronology protection is a fixed, nondynamical geometric datum that reduces the active gauge symmetry to $\mathrm{TDiff}_{\mathfrak{m}}$. Consequently, the associated Ward identity is weaker than the conservation law implied by full diffeomorphism invariance. When combined with the geometric Bianchi identity, it permits the \Moustafa{conditional} departure from ordinary-matter energy-momentum conservation expressed in Eq.~\eqref{MAGIC-conservation-equation}\Moustafa{, subject to the consistency Eq.~\eqref{consistencyExchange} required of the matter action.}

On an FLRW background, the chronology correction enters the Friedmann equations as an effective source with $w_{\mathrm{eff}}=-2$. The homogeneous Bianchi consistency condition then describes energy exchange with matter. The power-law profile considered in Sec.~\ref{sec:early-universe-example} illustrates how this exchange may occur during a finite early epoch and, for $n>3w$, decays faster than the ordinary perfect-fluid contribution while leaving a finite enhancement of the ordinary fluid's comoving energy.

We have also introduced a local scalar representation of \MAGIC{}, which clarifies the perturbative content of the homogeneous branch. Locally, \MAGIC{} reduces to the $w_{\mathrm{eff}}=-2$ scalar branch of generalized unimodular gravity. The \Moustafa{local, quadratic,} matter-free, spatially flat FLRW spectrum \Moustafa{on the non-compact $C>0$, $\Lambda=0$ branch} contains the two tensor polarizations of GR and one scalar mode, with no local kinetic mode in the multiplier or transverse projective sectors. The tensor sector remains healthy, and the scalar has a positive kinetic term, but $c_s^2=-2$ produces a short-wavelength gradient instability. \Moustafa{A viable cosmological application therefore requires stabilization.} Stabilizing the scalar without losing the timelike, closed projective mode is the natural next step toward a phenomenological completion.

These results turn the chronology-protection proposal into a concrete geometric construction and also identify several directions for future work.
First, Sec.~\ref{sec:perturbations} establishes only the local quadratic spectrum on the homogeneous branch. A background-independent Hamiltonian analysis is therefore needed to classify possible global or branch-dependent sectors. 
Second, Sec.~\ref{sec:foliation-regularity} shows that the spatial gradient of $m$ supplies a geometric acceleration term that can alter the focusing of the preferred congruence. The present analysis does not, however, establish the absence of caustics. Determining whether preventing metric degeneracy necessarily produces caustics or other singularities in the preferred congruence remains an open question.

The final direction returns to the black-hole motivation discussed in the introduction. The existence and structure of black-hole solutions in \MAGIC{} remain to be determined. In particular, the behavior of the renormalized energy-momentum tensor near the inner horizon of the Kerr black hole motivates asking whether the chronology-protecting sector modifies inner-horizon geometry, and if so, whether such modifications are observationally distinguishable from GR.

 \begin{acknowledgments}
We thank Walter Arata and Luke Martin for reading a draft of this paper and providing useful comments. We are also grateful to an anonymous referee for a careful and constructive report that substantially improved the manuscript.
 \end{acknowledgments}

\appendix

\section{Covariant differentiation of tensor densities}\label{App1}

We briefly review the covariant differentiation of tensor densities. For more comprehensive treatments, see Refs.~\cite{Weinberg:1972,IntegrationInGR}. 

A $(1,1)$-tensor density $\boldsymbol{\mathfrak{T}}$ of weight $w$ is a geometric object whose components transform under a general coordinate transformation $x^\mu\rightarrow x^{\mu'}$ as
\begin{align}
    {\mathfrak{T}^{  \mu '}}_{  \nu '} = J^{-w}   \frac{\partial x^{\mu'}}{\partial x^{\mu}} \frac{\partial x^{\nu}}{\partial x^{\nu'}}  {\mathfrak{T}^{  \mu }}_{  \nu },
\end{align}
where $J:= \det{\mathbf{J}}:= \det{\left(\partial x^{\mu'}/\partial x^{\nu} \right)}$ is the Jacobian determinant. Ordinary tensors correspond to $w=0$.

A covariant derivative of a tensor density is defined so that, if $\boldsymbol{\mathfrak{T}}$ is a $(k,l)$-tensor density of weight $w$, then $\boldsymbol{\nabla \mathfrak{T}}$ is a $(k,l+1)$-tensor density of the \textit{same} weight. Consider first a scalar density $\mathfrak{S}$ satisfying
\begin{align}
  \mathfrak{S} '=J^{-w} \mathfrak{S}.
\end{align}
Unlike in the scalar-field case, $\partial_\mu \mathfrak{S}$ fails to transform covariantly. Instead,
\begin{align}
    \partial_{  \mu '} \mathfrak{S} ' &=  \frac{\partial x^\mu}{\partial x^{\mu'}}\partial_\mu  \left(J^{-w} \mathfrak{S} \right) \notag\\
    & = J^{-w}   \frac{\partial x^\mu}{\partial x^{\mu'}} \partial_\mu \mathfrak{S} - w   \frac{\partial x^\mu}{\partial x^{\mu'}} \left(\partial  _\mu  \ln{J}\right) J^{-w} \mathfrak{S}.\label{PartialDerivativeOfDensity}
\end{align}
The extra term proportional to $w$ spoils covariance, an issue one does not face when $w=0$. This can be corrected by defining
\begin{align}\label{AA5}
    \nabla_\mu \mathfrak{S}:=\partial_\mu \mathfrak{S} -w \Gamma  ^\nu_{  \mu\nu } \mathfrak{S}.
\end{align}
The non-covariant transformation of $-w \Gamma^\nu_{\mu\nu} \mathfrak{S}$ exactly cancels the non-covariant part of Eq.~\eqref{PartialDerivativeOfDensity}. Thus,
\begin{align}
    \nabla_{  \mu '} \mathfrak{S} ' = J^{-w}   \frac{\partial x^\mu}{\partial x^{\mu'}} \nabla_\mu \mathfrak{S}.
\end{align}

To verify this, recall that the affine connection transforms as
\begin{align}\label{Transformation of Gamma}
\Gamma^{  \rho '}_{  \mu '  \nu '}=  \frac{\partial x^{\rho'}}{\partial x^{\rho}} \frac{\partial x^{\mu}}{\partial x^{\mu'}} \frac{\partial x^{\nu}}{\partial x^{\nu'}} \Gamma^\rho_{  \mu\nu }-   \frac{\partial x^{\mu}}{\partial x^{\mu'}} \frac{\partial x^{\nu}}{\partial x^{\nu'}} \frac{\partial^2 x^{\rho'}}{\partial x^\mu \partial x^{\nu}}.
\end{align}
Contracting $\rho'$ with $\nu'$ gives
\begin{align}
\Gamma^{  \nu '}_{  \mu '  \nu '}
&=   \frac{\partial x^{\mu}}{\partial x^{\mu'}} \Gamma^\nu_{  \mu\nu }
-   \frac{\partial x^{\mu}}{\partial x^{\mu'}}
\frac{\partial x^{\nu}}{\partial x^{\nu'}}
\frac{\partial^2 x^{\nu'}}{\partial x^\mu \partial x^{\nu}}  \notag \\
&=   \frac{\partial x^{\mu}}{\partial x^{\mu'}} \Gamma^\nu_{  \mu\nu }
-   \frac{\partial x^{\mu}}{\partial x^{\mu'}}
 \left(\mathbf{J}^{-1}\right)  ^\nu{}_{  \nu '}
\partial_\mu \mathbf{J}^{\nu'}{}_{\nu}  \notag \\
&=   \frac{\partial x^{\mu}}{\partial x^{\mu'}} \Gamma^\nu_{  \mu\nu }
-   \frac{\partial x^{\mu}}{\partial x^{\mu'}}
 \mathrm{Tr}\left(\mathbf{J}^{-1}\partial  _\mu  \mathbf{J}\right).
\end{align}
Using the identity
\begin{align}\label{determinant trace identity}
    \partial  _\mu  \ln{\left(\det{\mathbf{M}}\right)}=\mathrm{Tr}\left(\mathbf{M}^{-1}\partial  _\mu  \mathbf{M}\right),
\end{align}
where $\mathbf{M}$ is a square matrix, one obtains
\begin{align}\label{ContractedConnection}
\Gamma^{  \nu '}_{  \mu '  \nu '} =   \frac{\partial x^{\mu}}{\partial x^{\mu'}} \Gamma^\nu_{  \mu\nu } -    \frac{\partial x^{\mu}}{\partial x^{\mu'}} \partial_\mu \ln{J}.
\end{align}
Together, Eqs.~\eqref{PartialDerivativeOfDensity} and~\eqref{ContractedConnection} ensure that the expression given by Eq.~\eqref{A5} transforms covariantly.

In the same way, for $(1,1)$-tensor densities, we define
 \begin{align}\label{A7}
    \nabla  _\rho  {\mathfrak{T}^{  \mu }}_{  \nu }:= \partial  _\rho  {\mathfrak{T}^{  \mu }}_{  \nu } +   {\Gamma^\mu}_{\rho\sigma}   {\mathfrak{T}^{\sigma}} _{  \nu } -   {\Gamma^\sigma}_{\rho\nu}   {\mathfrak{T}^{\mu}} _{  \sigma }   - w {\Gamma^\sigma}_{\rho\sigma}{} \mathfrak{T}^{\mu}_{  \nu },
\end{align}
with the transformation rule
\begin{align}
    \nabla_{  \rho '} {\mathfrak{T}^{  \mu '}}_{  \nu '}
    =J^{-w}
      \frac{\partial x^\rho}{\partial x^{\rho'}}
    \frac{\partial x^{\mu'}}{\partial x^\mu}
    \frac{\partial x^\nu}{\partial x^{\nu'}}
    \nabla_\rho  {\mathfrak{T}^{  \mu }}_{  \nu }.
\end{align}
The standard covariant derivative is recovered for $w=0$. These results depend only on the transformation law of $\Gamma^\rho_{\mu\nu}$, irrespective of whether it is Levi--Civita or a more general affine connection.

Although we considered $(1,1)$-tensor densities for simplicity, all of the results extend straightforwardly to arbitrary type $(k,l)$.

\section{Variation of the metric--affine GR action with respect to the connection}\label{App2}
Variation of the metric--affine GR action with respect to the connection gives
\begin{align}\label{eq:26}
    \delta_\Gamma S_0  = \frac{1}{2} \int \mathrm{d}^4x \sqrt{-g} g^{\mu\nu} \delta R_{\mu\nu}[\mathbf{\Gamma}].
\end{align}
Using the \textit{Palatini identity},
\begin{align}\label{PalatiniIdentity}
    \delta R_{\mu\nu} \left(\Gamma\right) = \nabla  _\rho  \delta {\Gamma  ^\rho }_{\nu\mu} - \nabla_\nu \delta {\Gamma  ^\rho }_{  \rho \mu} - {T  ^\sigma }_{\nu  \rho } \delta {\Gamma  ^\rho }_{  \sigma \mu},
\end{align}
where ${T^\rho}_{\mu\nu}$ is the torsion tensor, and using the product rule, Eq.~\eqref{eq:26} becomes
\begin{align}\label{eq:30}
     \delta_\Gamma S_0  &= \frac{1}{2} \int \mathrm{d}^4x  \Big[ \nabla  _\rho  \left(\sqrt{-g} g^{\mu\nu}  \delta {\Gamma  ^\rho }_{\nu\mu}\right) \notag \\
     &- \nabla  _\rho  \left(\sqrt{-g} g^{\mu\nu}\right) \delta {\Gamma  ^\rho }_{\nu\mu} - \nabla_\nu \left(\sqrt{-g} g^{\mu\nu}  \delta {\Gamma  ^\rho }_{  \rho \mu}\right) \notag \\
     &+ \nabla_\nu \left(\sqrt{-g} g^{\mu\nu}\right) \delta {\Gamma  ^\rho }_{  \rho \mu} + \sqrt{-g} g^{\mu\nu}{T  ^\sigma }_{  \rho \nu } \delta {\Gamma  ^\rho }_{  \sigma \mu}\Big].
\end{align}

Because torsion is generally nonzero, surface terms must be treated carefully. Using Eq.~\eqref{A7} for a vector density $\mathfrak{U}^\mu$ of weight $w=1$, one has
\begin{align}
    \nabla_\mu \mathfrak{U}^\nu = \partial_\mu \mathfrak{U}^\nu + \Gamma^\nu_{\mu\sigma} \mathfrak{U}^\sigma -\Gamma^\sigma_{\mu\sigma} \mathfrak{U}^\nu.
\end{align}
Contracting $\mu$ with $\nu$,
\begin{align}
    \nabla_\mu \mathfrak{U}^\mu &= \partial_\mu \mathfrak{U}^\mu + \Gamma^\mu_{\mu\sigma} \mathfrak{U}^\sigma - \Gamma^\sigma_{\mu\sigma} \mathfrak{U}^\mu \notag \\
    &= \partial_\mu \mathfrak{U}^\mu + \Gamma^\mu_{\mu\sigma} \mathfrak{U}^\sigma - \Gamma^\mu_{\sigma\mu} \mathfrak{U}^\sigma \notag \\ &=\partial_\mu \mathfrak{U}^\mu + {T^\mu}_{\mu\sigma} \mathfrak{U}^\sigma.
\end{align}
Integration over a region $\mathcal{M}$ gives a \textit{generalized divergence theorem}
\begin{align}\label{byParts}
    \int_{\mathcal{M}} \mathrm{d}^4x \nabla_\mu \mathfrak{U}^\mu = \int_{\partial\mathcal{M}} \mathrm{d}^3y ~ n_\mu \mathfrak{U}^\mu + \int_{\mathcal{M}} \mathrm{d}^4x~ {T^\mu}_{\mu\sigma} \mathfrak{U}^\sigma,
\end{align}
where $n_\mu$ is the outward unit one-form normal to the boundary $\partial\mathcal{M}$.

With $\mathfrak{U}^\rho := \sqrt{-g} g^{\mu\nu}\delta {\Gamma^\rho}_{\nu\mu}$ and $\delta {\Gamma^\rho}_{\mu\nu}\big|_{\partial\mathcal{M}}=0$, we use Eq.~\eqref{byParts} to simplify Eq.~\eqref{eq:30}:
\begin{align}\label{GRGammaApp}
       \frac{\delta S_0 }{\delta {\Gamma^\rho}_{\mu\nu}} &= \frac{\sqrt{-g}}{2} \Big[ -{T  ^\sigma }_{  \sigma\rho } g^{\mu\nu} + {T  ^\sigma }_{  \sigma\lambda } g^{  \lambda \nu} \delta^\mu  _\rho  + {T^\mu}_{  \rho\sigma } g^{  \sigma \nu} \notag \\
     & - \frac{1}{\sqrt{-g}}\nabla  _\rho  \left(\sqrt{-g} g^{\mu\nu}\right) +\frac{1}{\sqrt{-g}}\nabla  _\sigma  \left(\sqrt{-g} g^{  \sigma \nu}\right) \delta^\mu  _\rho   \Big],
\end{align}
where surface terms have been dropped only after accounting for torsion.

\section{Variation of the MAGIC action with respect to the metric} \label{App4}
Variation of the \MAGIC{} action, Eq.~\eqref{Action}, with respect to the metric while holding $\mathbf{\Gamma}$ fixed gives
\begin{align}
    & 2\delta_{\mathrm{g}} S_{\mathrm{\MAGIC{}}} = 2\delta_{\mathrm{g}} S_0  + 2\int \mathrm{d}^4x ~\delta_{\mathrm{g}} \left(\sqrt{-g}\right) \xi^{\mu\nu} \partial_{[\mu}\Gamma^{  \rho }_{\nu]  \rho } \notag \\
    &+ \int \mathrm{d}^4x ~ \delta_{\mathrm{g}} \left(\frac{\lambda}{\sqrt{-g}^3}\right) \left[ \mathfrak{g}^{\mu\nu} \left(\nabla_\mu \sqrt{-g}\right) \left(\nabla_\nu \sqrt{-g}\right) + \mathfrak{m}^2 \right] \notag \\
    &+ \int \mathrm{d}^4x ~ \frac{\lambda}{\sqrt{-g}^3} \delta_{\mathrm{g}} \left[ \mathfrak{g}^{\mu\nu} \left(\nabla_\mu \sqrt{-g}\right) \left(\nabla_\nu \sqrt{-g}\right) + \mathfrak{m}^2 \right].
\end{align}
The second and third terms vanish because of the constraints in Eqs.~\eqref{eq:density_constraint} and~\eqref{gradient constraint}, leaving
\begin{align}\label{7.17}
    2\delta_{\mathrm{g}} S_{\mathrm{\MAGIC{}}} = 2\delta_{\mathrm{g}} S_0  +\int \mathrm{d}^4x \,\Delta,
\end{align}
where
\begin{align}
    \Delta:= \frac{\lambda}{\sqrt{-g}^3} \delta_{\mathrm{g}} \left[ \mathfrak{g}^{\mu\nu} \left(\nabla_\mu \sqrt{-g}\right) \left(\nabla_\nu \sqrt{-g}\right) + \mathfrak{m}^2 \right].
\end{align}
Using the product rule,
\begin{align}\label{7.18}
    \Delta & = \frac{\lambda}{\sqrt{-g}^3} \Big[ \delta \mathfrak{g}^{\mu\nu} \left(\nabla_\mu \sqrt{-g}\right) \left(\nabla_\nu \sqrt{-g}\right) \notag\\
    &~~~~~~~~~~~~~~~~~~~~~~~~~~~~+2\mathfrak{g}^{\mu\nu} \left(\nabla_\mu \delta\sqrt{-g}\right) \left(\nabla_\nu \sqrt{-g}\right) \Big] \notag \\
    & = \frac{\lambda \delta \mathfrak{g}^{\mu\nu}}{\sqrt{-g}^3} \left(\nabla_\mu \sqrt{-g}\right) \left(\nabla_\nu \sqrt{-g}\right) \notag \\
    &~~~~~~~~~~~~~~~~~~~~~~~- 2\delta\sqrt{-g} \nabla_\mu \left[ \frac{\lambda \mathfrak{g}^{\mu\nu}}{\sqrt{-g}^3} \nabla_\nu \sqrt{-g} \right] \notag \\
    & ~~~~~+ \nabla_\mu \left[ \frac{2\lambda \mathfrak{g}^{\mu\nu}}{\sqrt{-g}^3} \left(\delta\sqrt{-g}\right) \left(\nabla_\nu \sqrt{-g}\right) \right].
\end{align}
But from Eq.~\eqref{InverseMetric1} and the identity
\begin{align}\label{D4}
    \delta \sqrt{-g}^n = -\frac{n}{2} \sqrt{-g}^n  g_{\mu\nu} \delta g^{\mu\nu},
\end{align}
we have
\begin{align}\label{D5}
    \delta \mathfrak{g}^{\mu\nu} = \delta g^{  \rho\sigma } \sqrt{-g}^2 \left( \delta^\mu  _\rho \delta^\nu_\sigma  - g^{\mu\nu} g_{  \rho\sigma } \right).
\end{align}
We then substitute Eqs.~\eqref{D4} and~\eqref{D5} into Eq.~\eqref{7.18} and use Eq.~\eqref{InverseMetric1} to replace $\mathfrak{g}^{\mu\nu}$ by $g^{\mu\nu}$,
\begin{align}
    \Delta & = \frac{\lambda \delta g^{\rho\sigma} \left( \delta^\mu_\rho \delta^\nu_\sigma - g^{\mu\nu} g_{\rho\sigma} \right)}{\sqrt{-g}} \left(\nabla_\mu \sqrt{-g}\right) \left(\nabla_\nu \sqrt{-g}\right)\notag \\
    & ~~~~~+ \left( \sqrt{-g}  g_{  \rho\sigma } \delta g^{  \rho\sigma }\right) \nabla_\mu \left[ \frac{\lambda g^{\mu\nu}}{\sqrt{-g}} \nabla_\nu \sqrt{-g} \right] \notag \\
    & ~~~~~- \nabla_\mu \left[ \lambda g^{\mu\nu} \left(g_{  \rho\sigma } \delta g^{  \rho\sigma }\right) \left(\nabla_\nu \sqrt{-g}\right) \right].
\end{align}
Using Eqs.~\eqref{nonMetrVector},~\eqref{WeylVector}, and~\eqref{7.13}, we obtain after some algebra
\begin{align}\label{C7}
    \Delta & = \sqrt{-g} \delta g^{  \mu\nu } \Big[ \lambda \partial_\mu \phi \partial_\nu \phi + \left(\lambda m^2 +\nabla  _\rho  \left( \lambda g^{  \rho\sigma } \partial  _\sigma  \phi\right)\right) g_{  \mu\nu } \Big]  \notag \\
    & ~~~- \nabla  _\rho  \left[\delta g^{  \mu\nu } \lambda \sqrt{-g} g^{  \rho\sigma } g_{  \mu\nu } \partial  _\sigma  \phi \right].
\end{align}
Inserting Eq.~\eqref{C7} into Eq.~\eqref{7.17}, using the generalized divergence theorem in Eq.~\eqref{byParts}, dropping surface terms and simplifying, one finds
\begin{align}
    &2\delta_{\mathrm{g}} S_{\mathrm{\MAGIC{}}} = 2\delta_{\mathrm{g}} S_0  +\int \mathrm{d}^4x \sqrt{-g} \delta g^{  \mu\nu }  \Big\{ \lambda \partial_\mu \phi \partial_\nu \phi \notag\\
    &+ \left[\lambda m^2 - \lambda {T  ^\kappa }_{  \kappa\rho } g^{  \rho\sigma } \partial  _\sigma  \phi +\nabla  _\rho  \left( \lambda g^{  \rho\sigma } \partial  _\sigma  \phi\right) \right] g_{  \mu\nu } \Big\}.
\end{align}
Simplifying further using Eqs.~\eqref{TorsionTrace} and~\eqref{7.13},
\begin{align}\label{D12}
    &2\delta_{\mathrm{g}} S_{\mathrm{\MAGIC{}}} = 2\delta_{\mathrm{g}} S_0  +\int \mathrm{d}^4x \sqrt{-g} \delta g^{  \mu\nu } \notag \\
    & \times \Bigg\{ \lambda \partial_\mu \phi \partial_\nu \phi + \left[\frac{7}{4} \lambda m^2 +\nabla  _\rho  \left( \lambda g^{  \rho\sigma } \partial  _\sigma  \phi\right) \right] g_{  \mu\nu } \Bigg\}.
\end{align}
From Eqs.~\eqref{RelationBetweenNewAndOldCovariantDerivative} and~\eqref{7.13}, the last term can be expressed in terms of the Levi--Civita connection, and ultimately just partial derivatives. We obtain
\begin{align}\label{D13}
    \nabla  _\rho  \left( \lambda g^{  \rho\sigma } \partial  _\sigma  \phi\right) &= \overset{(g)}{\nabla}  _\rho  \left( \lambda g^{  \rho\sigma } \partial  _\sigma  \phi\right) - \frac{1}{4}\left(\partial  _\rho  \phi\right) \left( \lambda g^{  \rho\sigma } \partial  _\sigma  \phi\right) \notag \\
    & = \overset{(g)}{\nabla}  _\rho  \left( \lambda g^{  \rho\sigma } \partial  _\sigma  \phi\right) + \frac{1}{4} \lambda m^2 \notag \\
    & = \frac{1}{\sqrt{-g}} \partial  _\rho  \left( \lambda \sqrt{-g} g^{  \rho\sigma } \partial  _\sigma  \phi\right) + \frac{1}{4} \lambda m^2.
\end{align}
Substituting Eq.~\eqref{D13} into Eq.~\eqref{D12} and reading off the functional derivative,
\begin{align}
    & \frac{\delta S_{\mathrm{\MAGIC{}}}}{\delta g^{\mu\nu}} = \frac{\delta S_0 }{\delta g^{\mu\nu}} + \frac{\sqrt{-g}}{2} \Big\{ \lambda \partial_\mu \phi \partial_\nu \phi \notag \\
    &+ \Big[2 \lambda m^2 + \frac{1}{\sqrt{-g}} \partial  _\rho  \left( \lambda \sqrt{-g} g^{  \rho\sigma } \partial  _\sigma  \phi\right) \Big] g_{  \mu\nu } \Big\},
\end{align}
with $\delta S_0 /\delta g^{\mu\nu}$ given by Eq.~\eqref{Einstein}. Setting $\delta S_{\mathrm{\MAGIC{}}}/\delta g^{\mu\nu}=0$, we obtain
\begin{align}
    G_{  \mu\nu }  & +  \left[\Lambda +2 \lambda m^2 + \frac{1}{\sqrt{-g}} \partial_\rho \left( \lambda \sqrt{-g} g^{\rho\sigma} \partial_\sigma \phi\right) \right]  g  _{\mu\nu}  \notag \\
    & + \lambda \partial_\mu \phi \partial_\nu \phi =  T_{\mu\nu}.
\end{align}
The third term in the square brackets vanishes on account of Eq.~\eqref{phiConservation}. Thus,
\begin{align}
  G_{\mu\nu} +\left(\Lambda + 2 \lambda m^2  \right) g_{\mu\nu} + \lambda \partial_\mu \phi \partial_\nu \phi =  T_{  \mu\nu }.
\end{align}

\bibliography{apssamp}% Produces the bibliography via BibTeX.

\end{document}